\documentclass[final,twoside,11pt]{entics} 
\usepackage{enticsmacro}
\usepackage{epic,eepic}
\usepackage{pict2e}
\usepackage{graphicx}
\usepackage{tikz}
\usepackage{proof}
\usepackage{datetime}
\usepackage{url}
\newcommand{\calA}{\mathcal{A}}
\newcommand{\calC}{\mathcal{C}}
\newcommand{\calP}{\mathcal{P}}
\newcommand{\CA}{\PA}
\newcommand{\PA}{\mathcal{P}_\calA}
\newcommand{\IA}{\mathcal{I}_\calA} 
\newcommand{\ILam}{\mathcal{I}_\Lambda} 
\newcommand{\id}{\mathit{id}}
\newcommand{\app}{\mathbf{app}}
\newcommand{\sem}[1]{[\![#1]\!]}
\newcommand{\BB}{\mathbf{B}}
\newcommand{\CC}{\mathbf{C}}
\newcommand{\II}{\mathbf{I}}

\newcommand{\KK}{\mathbf{K}}
\newcommand{\WW}{\mathbf{W}}
\newcommand{\BCI}{\mathbf{BCI}}
\newcommand{\BCpmI}{\mathbf{BC^\pm{}I}}
\newcommand{\BIdot}{\mathbf{BI}(\_)^\bullet}
\newcommand{\SK}{\mathbf{SK}}
\newcommand{\BCIWK}{\mathbf{BCIWK}}
\newcommand{\comment}[1]{}
\newcommand{\tmpcomment}[1]{}
\newcommand{\cox}{\mathit{cox}}
\definecolor{blue}{rgb}{0, .1, 1}
\definecolor{keywordcolor}{rgb}{1, 0, .7}
\definecolor{theoremcolor}{rgb}{.7, 0, .4}
\definecolor{bibcolor}{rgb}{.6, .2, .8}
\definecolor{syntaxcolor}{rgb}{.2, .4, 1}
\definecolor{semanticscolor}{rgb}{.6, .2, .8}
\definecolor{appendixcolor}{rgb}{.8, .4, .2}
\definecolor{remarkcolor}{rgb}{.8, .4, 0}
\definecolor{definitioncolor}{rgb}{.2, .4, .1}

\def\red{\color{red}}

\def\keywordcolor{\color{black}}

\def\semcolor{\color{black}}
\def\remarkcolor{\color{remarkcolor}}

\newcommand{\kc}{\keywordcolor}

\newcommand{\lamnode}{\color[rgb]{1,0,0}\makebox(0,0){\huge$\bullet$}\color[rgb]{0,0,0}\makebox(0,0){\tiny$\lambda$}}

\newcommand{\appnode}{\color[rgb]{0,.6,.6}\makebox(0,0){\huge$\bullet$}\color[rgb]{0,0,0}\makebox(0,0){\tiny$@$}}

\newcommand{\tinylamnode}{\color[rgb]{1,0,0}\makebox(0,0){$\bullet$}}

\newcommand{\tinyappnode}{\color[rgb]{0,.6,.6}\makebox(0,0){$\bullet$}}

\newdimen\cboxwidth
\cboxwidth=\unitlength
\newdimen\cboxheight
\cboxheight=\unitlength
\newcommand{\rgbfbox}[8]{%
\cboxwidth=\unitlength
\cboxheight=\unitlength
\multiply \cboxwidth by #3
\multiply \cboxheight by #4
{\color[rgb]{#5,#6,#7}\put(#1,#2){\rule{\cboxwidth}{\cboxheight}}}
\put(#1,#2){\framebox(#3,#4){#8}}
\cboxwidth=\unitlength
\cboxheight=\unitlength
}
\newcommand{\rgbbox}[8]{%
\cboxwidth=\unitlength
\cboxheight=\unitlength
\multiply \cboxwidth by #3
\multiply \cboxheight by #4
{\color[rgb]{#5,#6,#7}\put(#1,#2){\rule{\cboxwidth}{\cboxheight}}}
\put(#1,#2){\makebox(#3,#4){#8}}
\cboxwidth=\unitlength
\cboxheight=\unitlength
}
\newsavebox{\boxp}
\savebox{\boxp}{%
\unitlength=.8pt
\begin{picture}(80,80)(0,20)\thicklines
\put(50,60){\circle{80}}
\put(50,100){\vector(-1,0){1}}
\put(11,69){\vector(-1,-4){1}}
\put(11,51){\vector(1,-4){1}}
\put(30,25.5){\vector(2,-1){1}}
\put(70.5,25.5){\vector(2,1){1}}
\put(89,51){\vector(1,4){1}}
\put(89,69){\vector(-1,4){1}}
\qbezier(15,80)(35,80)(35,80) \qbezier(65,80)(65,80)(85,80)
\qbezier(10,60)(35,60)(35,60) \qbezier(65,60)(65,60)(90,60)
\qbezier(15,40)(35,40)(35,40) \qbezier(65,40)(65,40)(85,40)
\put(30,80){\vector(1,0){1}}
\put(75,80){\vector(1,0){1}}
\put(30,60){\vector(1,0){1}}
\put(75,60){\vector(1,0){1}}
\put(30,40){\vector(1,0){1}}
\put(75,40){\vector(1,0){1}}
\put(35,35){\framebox(30,50){}}
\put(50,20){\lamnode}
\put(15,80){\lamnode}
\put(10,60){\lamnode}
\put(15,40){\lamnode}
\put(85,80){\appnode}
\put(90,60){\appnode}
\put(85,40){\appnode}
\end{picture}
}

\newsavebox{\hbraid}
\savebox{\hbraid}{%
\unitlength=.8pt
\begin{picture}(20,20)(0,0)\thicklines
\qbezier(0,20)(8,20)(10,10)\qbezier(10,10)(12,0)(20,0)
\rgbbox{6}{6}{8}{8}{1}{1}{1}{}%
\qbezier(0,0)(8,0)(10,10)\qbezier(10,10)(12,20)(20,20)
\end{picture}}

\newsavebox{\redhbraid}
\savebox{\redhbraid}{%
\unitlength=.8pt
\begin{picture}(20,20)(0,0)\thicklines
{\red\qbezier(0,20)(8,20)(10,10)\qbezier(10,10)(12,0)(20,0)}%
\rgbbox{6}{6}{8}{8}{1}{1}{1}{}%
\qbezier(0,0)(8,0)(10,10)\qbezier(10,10)(12,20)(20,20)
\end{picture}}

\newsavebox{\hbraidInv}
\savebox{\hbraidInv}{%
\unitlength=.8pt
\begin{picture}(20,20)(0,0)\thicklines
\qbezier(0,0)(8,0)(10,10)\qbezier(10,10)(12,20)(20,20)
\rgbbox{6}{6}{8}{8}{1}{1}{1}{}%
\qbezier(0,20)(8,20)(10,10)\qbezier(10,10)(12,0)(20,0)
\end{picture}}

\newcommand{\myline}[4]{\qbezier(#1,#2)(#1,#2)(#3,#4)}

\settimeformat{ampmtime}

\def\conf{MFPS 2022} 	
\volume{1}			
\def\volu{1}			
\def\papno{5}			
\def\mydoi{{\normalshape \href{https://doi.org/10.46298/entics.10338}{10.46298/entics.10338}}}
\def\copynames{M. Hasegawa} 

\def\runtitle{Internal Operads of Combinatory Algebras}

\def\lastname{Hasegawa}
\begin{document}
\begin{frontmatter}
  \title{The Internal Operads of Combinatory Algebras} 
  \author{Masahito Hasegawa\thanksref{ALL}\thanksref{myemail}}	
  \address{Research Institute for Mathematical Sciences\\ Kyoto University\\			
    Kyoto, Japan}  							
  \thanks[ALL]{I thank Haruka Tomita for 
  stimulating discussions, and the anonymous reviewers for 
  comments. This work was supported by JSPS KAKENHI Grant
No. JP18K11165, JP21K11753 and JST ERATO Grant No. JPMJER1603.}   
   \thanks[myemail]{Email: \href{mailto:hassei@kurims.kyoto-u.ac.jp} {\texttt{\normalshape
        hassei@kurims.kyoto-u.ac.jp}}} 
\begin{abstract} 
We argue that {\em operads} provide a general framework for dealing with 
polynomials and combinatory completeness of {\em combinatory algebras},
including the classical $\mathbf{SK}$-algebras, linear $\mathbf{BCI}$-algebras, 
planar $\mathbf{BI}(\_)^\bullet$-algebras as well as the braided 
$\mathbf{BC^\pm I}$-algebras. We show that every extensional combinatory algebra
gives rise to a canonical closed operad, which we shall call the {\em internal operad}
of the combinatory algebra. The internal operad construction gives a left adjoint to
the forgetful functor from closed operads to extensional combinatory algebras.
As a by-product, we derive extensionality axioms for
the classes of combinatory algebras mentioned above.
\end{abstract}
\end{frontmatter}
\section{Introduction}\label{intro}

{\em Combinatory algebras} \cite{Bar84,HS08} are fundamental in several areas of 
theory of computation. They can be thought as models of the
$\lambda$-calculus, in which the $\lambda$-abstraction is not a primitive ingredient
but a derived construct.
This paper addresses a seemingly naive and easy-to-answer question 
on this ability of modelling $\lambda$-abstractions in combinatory algebras:
what are the correct interpretations of variables?
For the classical (cartesian) combinatory algebras, our approach basically agrees with
that of Hyland \cite{Hyl17}. However, our work is motivated by non-classical variants of combinatory algebras, especially by a difficulty in formulating the braided
combinatory algebras along the line of our previous work \cite{Has22}.
Technically, we build our framework on top
of the case of planar combinatory algebras \cite{Tom21,Tom22}.

\subsection{Polynomials and Combinatory Completeness}

Recall that an (total) {\em\keywordcolor   applicative structure} 
(also called a {\em magma}) $\semcolor(\calA,\cdot)$
is a set $\semcolor \mathcal{A}$ equipped with
a binary function $\semcolor (\_)\cdot(\_):\calA\times\calA\rightarrow\calA$ called
{\em\keywordcolor   application}.
In this paper we only deal with total applicative structures, i.e., applications are always 
defined. As is customary, applications are assumed to be left associative, and the infix $\cdot$ is often omitted. 

We are mainly 
interested in applicative structures which can model the
{\em\keywordcolor  $\lambda$-calculus}.
So we are to handle {\em variables} and {\em abstractions}.
Usually, we proceed as follows.

\begin{enumerate}
\item 
Introduce {\em\keywordcolor   polynomials} $\semcolor \mathcal{P}[x_1,\dots,x_n]$
on $\semcolor (\calA,\cdot)$, which are 
generated by variables $\semcolor x_1,\dots,x_n$ and elements of $\semcolor \calA$ 
using  the application $\semcolor \cdot$.  
\item 
We say that $\semcolor (\calA,\cdot)$ is {\em\keywordcolor   combinatory complete} if, 
for any $\semcolor p\in\mathcal{P}[\Gamma,x]$
there exists $\semcolor \lambda^*x. p\in\mathcal{P}[\Gamma]$ 
such that $\semcolor (\lambda^*x.p)\cdot q=p[q/x]$ holds.
We say it is {\em\keywordcolor extensional} when such $\semcolor \lambda^*x. p$ is 
unique (equivalently: $\lambda^*x.(p\cdot x)=p$ with no free $x$ in $p$).
\end{enumerate} 

In this paper, by a {\em\kc combinatory algebra} we mean
 a combinatory complete applicative structure.
Note the ambiguity in the notion of polynomials;
by altering the definition of polynomials, we get different notions of 
combinatory algebras. 

\begin{remark}
The extensionality above is not quite a standard one found in e.g. \cite{Bar84,HS08}, where
an applicative structure $\calA$ is called extensional when $(\forall x\in\calA~a\cdot x=b\cdot x)$ implies $a=b$. This traditional extensionality is too strong for our purpose; in particular, interesting  braided combinatory algebras 
cannot be extensional in the traditional sense, as it enforces braids with the same underlying permutation to be identified.
In contrast, our extensionality (or the $\eta$-rule)
says that
 $a\cdot x=b\cdot x$ in $\calP[\Gamma,x]$ implies $a=b$ for $a,b\in\calP[\Gamma]$, which heavily depends on the notion of polynomials. 
 \end{remark}

\subsection{Semi-closed Operads and Combinatory Algebras}

Thanks to combinatory completeness, a combinatory algebra gives rise to a model of the
$\lambda$-calculus:
\begin{enumerate}
\item Firstly, polynomials $\semcolor \mathcal{P}[x_1,\dots,x_n]$ are thought as the set $\semcolor \calP(n)$
of 
$n$-ary operators on $\semcolor \calA$. 
\item The family $\semcolor \{\calP(n)\}_{n\in\mathbf{N}}$ with suitable notion of composition 
determines an {\em\keywordcolor operad} (one-object multicategory) $\semcolor \calP$ with 
$\semcolor \calP(0)=\calA$.
Depending on the definition of polynomials, the operad can be {\em\kc planar}, {\em\kc symmetric}, {\em\kc braided} or {\em\kc cartesian}.
\item Then, combinatory completeness says the operad $\semcolor \calP$ is
{\em\kc semi-closed} ({\em\kc closed} when extensional).
\item It is not hard to see that semi-closed (or closed) planar/symmetric/braided/cartesian operads are models of the planar/linear/braided/ordinary $\lambda$-calculus with $\beta$-
(or $\beta\eta$-)equality, where a term-in-context $x_1,\dots,x_n\vdash M$ is
interpreted as an element $\sem{x_1,\dots,x_n\vdash M}$ of $\calP(n)$.
\end{enumerate}
So the situation can be summarized as follows:
$$
\begin{array}{rcl}
\mbox{Constructing polynomials} & = & \mbox{Constructing operads}
\\
\mbox{Requiring combinatory completeness} &=& \mbox{Requiring (semi-)closedness}
\end{array}
$$
Conversely, 
(semi-)closed operads give a combinatory algebra, just by taking the $0$-ary operators:

\begin{itemize}
\item the planar case: when $\semcolor \calP$ is a semi-closed planar operad,  
$\semcolor \calP(0)$
is a {\em $\BIdot$-algebra} of Tomita \cite{Tom21};
\item the linear case: when $\semcolor \calP$ is a semi-closed symmetric operad, 
 $\semcolor \calP(0)$ 
is a {\em $\BCI$-algebra} \cite{AL05,Hos07};
\item the braided case: when $\semcolor \calP$ is a semi-closed braided operad, 
 $\semcolor \calP(0)$ 
is a {\em $\BCpmI$-algebra}, a braided variant of $\BCI$-algebras  
\cite{Has22}; and
\item the classical case: when $\semcolor \calP$ is a semi-closed cartesian operad,
$\semcolor \calP(0)$ is an {\em $\mathbf{SK}$-algebra} \cite{Hyl17}.
\end{itemize}
It remains to see how to construct polynomials, or more generally operads, 
on top of a combinatory algebra. 

\subsection{Taking Polynomials Seriously}

Often, polynomials of $\semcolor \calP[x_1,\dots,x_n]$ are identified with 
certain functions from $\semcolor \calA^n$ to $\semcolor \calA$. 
Many studies
on combinatory algebras employ this ``{\kc polynomials as functions}'' view
either explicitly or implicitly (e.g. formulating polynomials as formal expressions
while saying that two polynomials are equal when they express the same function).

There also are cases handling polynomials as a ``polynomial combinatory algebra'' in the algebraic manner \cite{Fre89,Sel02}, in which variables are  
taken as indeterminates. This approach allows a cleaner treatment of
abstractions where the problem of $\xi$-rule disappears \cite{Sel02}, and mathematically 
preferable than the  ``{\kc polynomials as functions}'' approach.
However, for the planar, linear and braided cases, polynomials do not form a combinatory algebra, and we cannot apply the same strategy.
 
Note that operads obtained in the ``{\kc polynomials as functions}'' way
are always {\kc well-pointed}:
the global section multi-functor into $\semcolor \mathbf{Set}$ is faithful. 
Hence many of the conventional approaches actually consider only well-pointed operads.
It still works reasonably well for the classical, linear and planar cases
(modulo the problem of $\xi$-rule).
However, the same cannot be applied to the {\kc braided} case: well-pointed
braided operads are always symmetric, hence the information on braids is lost.
Thus existing approaches are too restrictive.  
Is there an alternative way of constructing operads from combinatory algebras
which can cover the braided case?

\subsection{The Internal Operad Construction}

In this paper, 
we propose an alternative construction of operads from 
combinatory algebras: the {\em \kc internal operad} construction.
The key insight is that, instead of taking {\em external} functions as polynomials,
we construct an operad just by using the elements and structure of the combinatory algebra, hence {\em internally}.
As we will explain in Section \ref{sec:motivation}, its basic idea is rather simple and 
should be unsurprising for those familar with the $\lambda$-calculus: just to 
express a program with $m$ inputs and $n$ outputs by a closed $\lambda$-term 
of the form 
$\lambda f x_1\dots x_m.f\,M_1\dots M_n$
with no free $f$ in $M_i$'s. The novel finding is that it works in a wide class of 
combinatory algebras, in which we can characterise elements of $m$ inputs and 
$n$ outputs by an equation.
We show that the internal operad is the initial
one among the closed operads giving rise to the combinatory algebra. In other words, 
the internal operad construction is left adjoint to the functor
sending a closed operad $\calP$ to the extensional 
combinatory algebra $\calP(0)$.  

Moreover, the internal operad construction works not only for the planar, linear and
classical (non-linear) cases but also for the {\kc  braided} case.
This gives an answer to the difficulty of formalizing polynomials and combinatory completeness of braided combinatory algebras.

The main restriction of this approach is that the internal operad construction works only for {\em extensional} combinatory algebras. 
In fact, we can {\em design} extensionality axioms so that the 
internal operad construction works; this might be compared to Freyd's 
approach to extensionality \cite{Fre89} where he identifies axioms to
make the ``polynomial combinatory algebra'' construction satisfy the extensional principle.
The axioms obtained in this way are semantically motivated and (hopefully)
understandable. We present the resulting axiomatizations for the planar, linear, braided as well as the  classical cases.

\subsection{Related Work}

This work started with the question of how to formulate 
combinatory completeness of braided combinatory algebras,
which came from our previous work on the braided $\lambda$-calculus \cite{Has22}.
The notion of $\BCpmI$-algebras also comes from that work, though its axiomatization
was left open.

Hyland \cite{Hyl17} advocated the view that (classical) combinatory algebras 
are semi-closed cartesian operad. Our approach can be seen as generalization of 
his work to planar, linear, and braided settings. The main difference would be
that we put the planar -- the weakest but most general  -- case as the basic
setting, and develop other cases on top of it.

The planar combinatory algebras -- $\BIdot$-algebras and variations -- have
been studied by Tomita \cite{Tom21,Tom22} as the realizers for his 
non-symmetric realizability models. 

There are plenty of work on the graphical presentations of the $\lambda$-calculus;
while many focus on the graph-theoretic or combinatorial aspects,
 Zeilberger's work on linear/planar $\lambda$-terms and trivalent graphs 
 \cite{Zei16,Zei18} provide a more geometric perspective on the graphs,
which is closer to our approach.

Ikebuchi and Nakano's work on $\BB$-terms \cite{IN19} emphasizes the
role of composition and application of $\BB$ as basic constructs of 
their calculus of $\BB$-terms as forest of binary trees, which is very close to
our definition of internal operads; only the identity $\II$ and the internalization operator
$(\_)^\bullet$ are missing.

Some work on knotted graphs (including \cite{Kau89,Yet89}) identify
the ``Reidemeister-IV'' move, which is used in our axiomatizations of extensional
$\BCI$-algebras and $\BCpmI$-algebras.

\subsection{Organization of This Paper}
This paper is organized as follows. 
In Section \ref{sec:motivation}, we consider the combinatory algebras of closed
$\lambda$-terms as well as its graphical variants, and see how they give rise
to operads internally. 
In Section \ref{sec:planar}, we review Tomita's $\BIdot$-algebras from
the viewpoint of planar operads, and introduce extensional $\BIdot$-algebras.
In Section \ref{sec:internal}, we introduce internal operads of 
extensional $\BIdot$-algebras. 
Section \ref{sec:variations} is devoted to the cases of linear, braided and classical combinatory algebras, which are obtained by specializing the planar case with additional structures.
In Section \ref{sec:conclusion}, we conclude this paper by suggesting possible future work, including the preliminary observations on traced combinatory algebras.
For lack of space most proofs are omitted, though they all follow from plain equational reasoning. 
We assume that the reader is familiar with the basic concepts of the $\lambda$-calculus
and combinatory logic as found e.g., in \cite{HS08}. Brief summaries of 
the braid groups and 
braided operads used in this paper are given in Appendix  \ref{sec:braided-operads}.

\section{Motivating Internal Operads} \label{sec:motivation}

\subsection{The Planar, Linear, and Braided $\lambda$-calculi}
\label{subsec:lambda-calculi}

Let us summarize the fragments and variant of the $\lambda$-calculus to be 
discussed in this paper.
The {\em planar $\lambda$-calculus} is an untyped linear $\lambda$-calculus 
with no exchange, whose terms are given by the following rules.
$$
\infer[\mathrm{variable}]{x\vdash x}{}
~~~~~~
\infer[\mathrm{abstraction}]{\Gamma\vdash \lambda x.M}{\Gamma,x\vdash M}
~~~~~~
\infer[\mathrm{application}]{\Gamma,\Gamma'\vdash M\,N}{\Gamma\vdash M & \Gamma'\vdash N}
$$
It is easy to see that planar terms are closed under $\beta\eta$-conversion.
Typical planar terms include $\II=\lambda f.f$, $\BB=\lambda fxy.f\,(x\,y)$,
and $P^\bullet=\lambda f.f\,P$ for planar closed term $P$.

The {\em linear $\lambda$-calculus} has the rules for the planar $\lambda$-calculus
and the exchange rule:
$$
\infer[\mathrm{exchange}]
{x_{s(1)},x_{s(2)},\dots,x_{s(n)}\vdash M}
{x_1,x_2,\dots,x_n\vdash M & s:\mbox{permutation on $\{1,\dots,n\}$}}
$$
Non-planar linear terms include $\CC=\lambda fxy.f\,y\,x$.

The {\em braided $\lambda$-calculus} \cite{Has22} is a variant of the linear $\lambda$-calculus
in which every 
permutation/exchange of variables is realized by a braid. 
Thus, for a term $M$ with $n$ free variables and a braid $s$
with $n$ strands (which can be identified with the elements of the braid group $B_n$
as explained in Appendix \ref{sec:braided-operads}
below), we introduce a term $[s]M$ 
in which the 
free variables are permutated by $s$:
$$
\infer[\mathrm{braid}]
{x_{s(1)},x_{s(2)},\dots,x_{s(n)}\vdash [s]M}
{x_1,x_2,\dots,x_n\vdash M & s:\mbox{braid with $n$ strands}}
$$
For instance, there are infinitely many braided $\mathbf{C}$-combinators 
including
$$
\mathbf{C}^+ = \lambda fxy.
\left[
\begin{picture}(24,16)(-2,-2)
\thicklines
\qbezier(0,0)(5,0)(10,5)
\qbezier(10,5)(15,10)(20,10)
\qbezier(0,10)(5,10)(7,8)
\qbezier(13,2)(15,0)(20,0)
\qbezier(0,-10)(0,-10)(20,-10)
\end{picture}
\right]
(f\,y\,x)
~~\mbox{and}~~
\mathbf{C}^- = \lambda fxy.
\left[
\begin{picture}(24,16)(-2,-2)
\thicklines
\qbezier(0,10)(5,10)(10,5)
\qbezier(10,5)(15,0)(20,0)
\qbezier(0,0)(5,0)(7,2)
\qbezier(13,8)(15,10)(20,10)
\qbezier(0,-10)(0,-10)(20,-10)
\end{picture}
\right]
(f\,y\,x).
$$
The $\beta\eta$-equality on braided terms is less straightforward due to the
presence of braids;  see \cite{Has22} for details.

\subsection{Operads}

Recall that an {\keywordcolor (planar or non-symmetric) operad} \cite{Lei04}
$\semcolor \mathcal{P}$ is a family of sets $\semcolor (\mathcal{P}(n))_{n\in\mathbf{N}}$ equipped with 
\begin{itemize}
\item an {\keywordcolor identity} $\semcolor \id\in\mathcal{P}(1)$ and
\item
a {\keywordcolor composition} map
sending $\semcolor f_i\in\mathcal{P}(k_i)$ ($\semcolor 1\leq i\leq n$) and 
$\semcolor g\in\mathcal{P}(n)$ to the composite
$\semcolor g(f_1,\dots,f_n)
\in\mathcal{P}(k_1+k_2+\dots + k_n)$
\end{itemize}
which are subject to the unit law and associativity:
$$
\begin{array}{cc}
f(\id,\dots,\id) = f = \id(f)
\\
h(g_1(f_{11},\dots,f_{1j_1}),\dots,g_k(f_{k1},\dots,f_{kj_k}))
=
(h(g_1,\dots,g_n))(f_{11},\dots,f_{km_k})
\end{array}
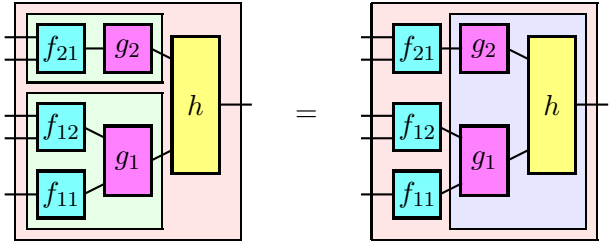
$$
\begin{center}
\unitlength=.85pt
\begin{picture}(110,105)(-5,-5)
\rgbfbox{0}{-10}{100}{105}{1}{.9}{.9}{}%
\rgbfbox{5}{60}{60}{30}{.9}{1}{.9}{}%
\rgbfbox{5}{-5}{60}{60}{.9}{1}{.9}{}%
\thicklines
\rgbfbox{70}{20}{20}{60}{1}{1}{.5}{$h$}%
\rgbfbox{40}{65}{20}{20}{1}{.5}{1}{$g_2$}%
\rgbfbox{40}{10}{20}{30}{1}{.5}{1}{$g_1$}%
\rgbfbox{10}{65}{20}{20}{.5}{1}{1}{$f_{21}$}%
\rgbfbox{10}{30}{20}{20}{.5}{1}{1}{$f_{12}$}%
\rgbfbox{10}{0}{20}{20}{.5}{1}{1}{$f_{11}$}%
\myline{-5}{80}{10}{80}
\myline{-5}{70}{10}{70}
\myline{-5}{45}{10}{45}
\myline{-5}{35}{10}{35}
\myline{-5}{10}{10}{10}
\myline{30}{75}{40}{75}
\myline{30}{40}{40}{35}
\myline{30}{10}{40}{15}
\myline{60}{25}{70}{30}
\myline{60}{75}{70}{70}
\myline{90}{50}{105}{50}
\end{picture}
\begin{picture}(40,105)
\put(20,50){\makebox(0,0){$=$}}
\end{picture}
\begin{picture}(110,105)(-5,-5)
\rgbfbox{0}{-10}{100}{105}{1}{.9}{.9}{}%
\rgbfbox{35}{-5}{60}{95}{.9}{.9}{1}{}%
\thicklines
\rgbfbox{70}{20}{20}{60}{1}{1}{.5}{$h$}%
\rgbfbox{40}{65}{20}{20}{1}{.5}{1}{$g_2$}%
\rgbfbox{40}{10}{20}{30}{1}{.5}{1}{$g_1$}%
\rgbfbox{10}{65}{20}{20}{.5}{1}{1}{$f_{21}$}%
\rgbfbox{10}{30}{20}{20}{.5}{1}{1}{$f_{12}$}%
\rgbfbox{10}{0}{20}{20}{.5}{1}{1}{$f_{11}$}%
\myline{-5}{80}{10}{80}
\myline{-5}{70}{10}{70}
\myline{-5}{45}{10}{45}
\myline{-5}{35}{10}{35}
\myline{-5}{10}{10}{10}
\myline{30}{75}{40}{75}
\myline{30}{40}{40}{35}
\myline{30}{10}{40}{15}
\myline{60}{25}{70}{30}
\myline{60}{75}{70}{70}
\myline{90}{50}{105}{50}
\end{picture}
\end{center}
$\semcolor \mathcal{P}(n)$ serves as the set of $n$-ary operators,
or polynomials with $n$ variables.

\subsection{(Semi-)Closed Operads and Combinatory Completeness}

From an applicative structure $\semcolor \calA$, we are to construct an
operad $\semcolor \mathcal{P}$ with $\semcolor \mathcal{P}(0)=\calA$
and an element $\semcolor \mathbf{app}\in\mathcal{P}(2)$ corresponding to the application $\semcolor\cdot$. 
That is, $\semcolor \app(a,b)=a\cdot b$ for $\semcolor a,b\in\calA$.

We say $\semcolor\calA$ is {\keywordcolor combinatory complete} with respect to $\semcolor\calP$ if,
for any $\semcolor p\in\mathcal{P}(n+1)$, there exists $\semcolor \lambda^*(p)\in\mathcal{P}(n)$
satisfying $\semcolor \mathbf{app}(\lambda^*(p),\id)=p$;
it is {\kc extensional} when such $\semcolor \lambda^*(p)$ is unique.

On the other hand, an operad $\semcolor \calP$ is {\kc semi-closed}
when there is  $\semcolor\mathbf{app}\in\calP(2)$ such that
for any $p\in\mathcal{P}(n+1)$,
there exists $\lambda^*(p)\in\mathcal{P}(n)$ satisfying
$\mathbf{app}(\lambda^*(p),\id)=p$, and 
{\kc closed} when such $\semcolor \lambda^*(p)$ is unique.
$$
\begin{array}{rcl}
\calP~\mbox{is semi-closed} 
& \Longleftrightarrow &
\calA\!=\!\calP(0)~\mbox{is combinatory complete with respect to}~\calP\\
\calP~\mbox{is closed} 
& \Longleftrightarrow &
\calA\!=\!\calP(0)~\mbox{is combinatory complete and extensional with respect to}~\calP
\end{array}
$$

\subsection{The Internal Operad of the $\lambda$-calculus}

The idea of the {\keywordcolor  internal operads} (and internal PRO(P)) 
is very simple if we look at the case of the combinatory algebra
of closed $\lambda$-terms, with its graphical interpretation.

We say that a closed $\lambda$-term is of {\keywordcolor arity} $\semcolor m\rightarrow n$
when it is $\beta\eta$-equal to a head normal form
$$\semcolor 
\lambda f x_1\dots x_m.f\,M_1\dots M_n~~~(\mbox{$f$ not free in $M_i$'s})
$$%
which can be regarded as a program with $m$ inputs and $n$ outputs, where
the head variable $\semcolor f$ serves as the (linearly-used) continuation or
the environment.
\comment{
If the reader is familiar with the polymorphic $\lambda$-calculus,  it might be
useful to notice that
such terms are inhabitants of the type
$$
\forall X.(B_1\!\rightarrow\!\cdots B_n\!\rightarrow\! X)\!\rightarrow A_1\!\rightarrow\!\cdots A_m\!\rightarrow\! X
\,\cong\,
(A_1\!\times\!\cdots\!\times\! A_m)\!\rightarrow\!(B_1\!\times\!\cdots\!\times\! B_n)$$
for regarding them as programs with $m$ inputs and $n$ outputs. 
The head variable $\semcolor f$ serves as the (linearly-used) continuation or
the environment.
}
There are closed terms which do not have an arity (e.g. $\lambda xy.y\,x$),
but we shall note that any closed term of the planar $\lambda$-calculus has an arity.
Examples of closed terms with arity include:
$$\semcolor 
\begin{array}{ccc}
\II=\lambda f.f:0\rightarrow 0 
&
~~~\BB=\lambda fxy.f\,(x\,y):2\rightarrow 1~~~
&
\CC=\lambda fxy.f\,y\,x:2\rightarrow 2
\\
\mathbf{S}=\lambda fxy.f\,y\,(x\,y):2\rightarrow 2
&
\mathbf{K}=\lambda fx.f:1\rightarrow 0
&
\mathbf{W}=\lambda fx.f\,x\,x:1\rightarrow 2
\\
\multicolumn{3}{c}{P^\bullet=\lambda f.f\,P:0\rightarrow 1~~(P~\mbox{closed term})}
\end{array}
$$%
Note that $M:m\rightarrow n$ implies $M:m+1\rightarrow n+1$ because 
we take the $\eta$-rule into account:
$$\semcolor 
\lambda f x_1\dots x_m.f\,M_1\dots M_n
\,=_\eta\,
\lambda f x_1\dots x_m x_{m+1}.f\,M_1\dots M_n\,x_{m+1}.
$$%
By letting $\semcolor \ILam(n)$ be the set
of ($\beta\eta$-equivalence classes of) closed terms of arity $\semcolor n\rightarrow 1$ and by appropriately defining the composition (with the identity $\II:1\rightarrow1$), 
we obtain a closed (cartesian) operad $\semcolor \ILam$, which we shall call
the {\em\keywordcolor  internal operad} of the $\lambda$-calculus.
The closed operad structure of $\ILam$  will be spelled out below; but before that, we
shall look at a graphical interpretation of terms with arity, which turns out to be
useful in describing the operad structure.

\subsection{The Internal Operad of the $\lambda$-calculus, Graphically}

We can interpret closed (linear) $\lambda$-terms as rooted trivalent graphs
with two kinds of nodes (the {\color[rgb]{1,0,0}lambda nodes} 
{\unitlength=.7pt\begin{picture}(10,10)\put(5,5){\lamnode}\end{picture}}
 and {\color[rgb]{0,.4,.4}application nodes}
{\unitlength=.7pt\begin{picture}(10,10)\put(5,5){\appnode}\end{picture}} 
) \cite{Zei16,Zei18} as shown in Figure \ref{fig:nodes}
\begin{figure}

\begin{minipage}[b]{0.25\linewidth}
\begin{center}
\unitlength=.7pt
\begin{picture}(120,60)(0,-5)
\thicklines
\put(10,50){\line(1,-1){20}} \put(15,45){\vector(1,-1){5}}
\put(30,30){\line(1,1){20}} \put(40,40){\vector(1,1){8}}
\put(30,30){\line(0,-1){25}}
\put(30,10){\vector(0,-1){3}}
\put(30,30){\lamnode}
\put(30,0){\makebox(0,0){\footnotesize$\lambda x.M$}}
\put(10,55){\makebox(0,0){\footnotesize$M$}}
\put(50,55){\makebox(0,0){\footnotesize$x$}}
\put(70,50){\line(1,-1){20}} \put(75,45){\vector(1,-1){5}}
\put(90,30){\line(1,1){20}} \put(105,45){\vector(-1,-1){5}}
\put(90,30){\line(0,-1){25}}
\put(90,10){\vector(0,-1){3}}
\put(90,30){\appnode}
\put(90,0){\makebox(0,0){\footnotesize$M\,N$}}
\put(70,55){\makebox(0,0){\footnotesize$M$}}
\put(110,55){\makebox(0,0){\footnotesize$N$}}
\end{picture}
\end{center}
\caption{The nodes}
\label{fig:nodes}
\end{minipage}
\begin{minipage}[b]{0.4\linewidth}
\begin{center}
\unitlength=.7pt
\begin{picture}(120,85)(0,-10)
\thicklines
\qbezier(0,60)(0,60)(20,40) \qbezier(40,60)(40,60)(20,40) 
\qbezier(20,40)(20,40)(20,20)
\qbezier(20,20)(20,20)(0,0) \qbezier(20,20)(20,20)(40,0)
\put(20,40){\lamnode}
\put(20,20){\appnode}
\put(20,25){\vector(0,-1){1}}
\put(11,49){\vector(1,-1){1}} \put(35,55){\vector(1,1){1}}
\put(5,5){\vector(-1,-1){1}} \put(29,11){\vector(-1,1){1}}
\put(-3,67){\makebox(0,0){\footnotesize$M$}}
\put(43,67){\makebox(0,0){\footnotesize$x$}}
\put(0,30){\makebox(0,0){\footnotesize$\lambda x.M$}}
\put(-3,-7){\makebox(0,0){\footnotesize$(\lambda x.M)\,N$}}
\put(43,-7){\makebox(0,0){\footnotesize$N$}}
\put(60,35){\makebox(0,0){$\stackrel{\beta}{=}$}}
\qbezier(80,60)(90,50)(90,30)\qbezier(90,30)(90,10)(80,0)
\qbezier(120,60)(110,50)(110,30)\qbezier(110,30)(110,10)(120,0)
\put(90,24){\vector(0,-1){1}}
\put(110,36){\vector(0,1){1}}
\put(75,67){\makebox(0,0){\footnotesize$M$}}
\put(120,67){\makebox(0,0){\footnotesize$x$}}
\put(75,-7){\makebox(0,0){\footnotesize$M$}}
\put(120,-7){\makebox(0,0){\footnotesize$N$}}
\end{picture}
\begin{picture}(50,50)(0,-5)
\end{picture}
\begin{picture}(100,65)(0,-10)
\thicklines
\qbezier(20,65)(20,60)(20,50)
\put(20,30){\circle{40}}
\qbezier(20,10)(20,0)(20,-5)
\put(20,50){\appnode}
\put(20,10){\lamnode}
\put(20,70){\makebox(0,0){\footnotesize$M$}}
\put(20,-10){\makebox(0,0){\footnotesize$\lambda x.M\,x$}}
\put(-5,50){\makebox(0,0){\footnotesize$M\,x$}}
\put(35,50){\makebox(0,0){\footnotesize$x$}}
\put(60,35){\makebox(0,0){$\stackrel{\eta}{=}$}}
\qbezier(90,60)(90,60)(90,0)
\put(0,24){\vector(0,-1){1}}
\put(40,36){\vector(0,1){1}}
\put(90,24){\vector(0,-1){1}}
\put(20,55){\vector(0,-1){1}}
\put(20,-5){\vector(0,-1){1}}
\put(90,70){\makebox(0,0){\footnotesize$M$}}
\put(90,-10){\makebox(0,0){\footnotesize$M$}}
\end{picture}
\end{center}
\caption{$\beta\eta$-rules}
\label{fig:beta-eta}
\end{minipage}
\begin{minipage}[b]{0.35\linewidth}
\begin{center}
\unitlength=.7pt
\begin{picture}(100,120)(0,20)
\thicklines
\put(50,90){\circle{100}}
\rgbfbox{20}{65}{60}{50}{1}{1}{0}{}

\qbezier(5,110)(5,110)(20,110) \qbezier(80,110)(95,110)(95,110)
\qbezier(0,90)(0,90)(20,90) \qbezier(80,90)(100,90)(100,90)
\qbezier(5,70)(5,70)(20,70) \qbezier(80,70)(95,70)(95,70)
\put(5,110){\vector(1,0){15}}
\put(0,90){\vector(1,0){20}}
\put(5,70){\vector(1,0){15}}

\put(92,110){\vector(1,0){1}}
\put(92,90){\vector(1,0){1}}
\put(92,70){\vector(1,0){1}}
\put(50,40){\vector(0,-1){20}}
\put(50,140){\vector(-1,0){1}}
\put(2,100){\vector(-1,-4){1}}
\put(2,80){\vector(1,-4){1}}
\put(98,100){\vector(-1,4){1}}
\put(98,80){\vector(1,4){1}}
\put(20,50){\vector(1,-1){1}}
\put(80,50){\vector(1,1){1}}
\put(5,110){\lamnode}
\put(0,90){\lamnode}
\put(5,70){\lamnode}
\put(50,40){\lamnode}
\put(95,110){\appnode}
\put(100,90){\appnode}
\put(95,70){\appnode}
\put(-10,90){\makebox(0,0){$\Bigg\{$}}
\put(110,90){\makebox(0,0){$\Bigg\}$}}
\put(-32,90){\makebox(0,0){$m$~~~s}}
\put(-27,91){\lamnode}
\put(130,90){\makebox(0,0){$n$~~~s}}
\put(132,91){\appnode}
\end{picture}
\end{center}
\caption{Graphs of arity $m\rightarrow n$}
\label{fig:m-to-n}
\end{minipage}

\end{figure}
where the annotations show the correspondence to the linear $\lambda$-terms.
They are 
subject to the $\beta\eta$-rules given in Figure \ref{fig:beta-eta}.

We are interested in the graphs (modulo $\beta\eta$-rules) of arity $m\rightarrow n$
as depicted in Figure \ref{fig:m-to-n},
which are $\semcolor\lambda fx_1\dots x_m.f\,M_1\,\dots M_n$ in the 
$\lambda$-calculus.
The most basic examples of such graphs with arity are:
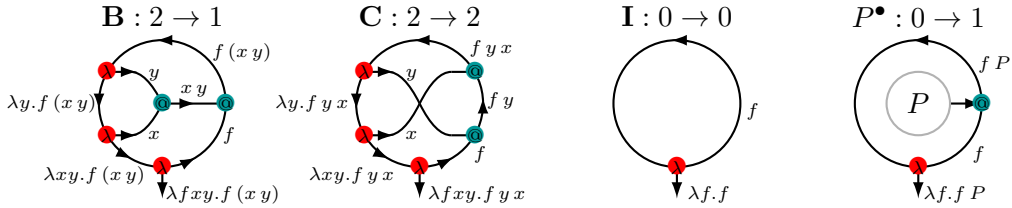
\begin{center}
\unitlength=.6pt
\begin{picture}(100,120)
\put(50,115){\makebox(0,0){$\BB:2\rightarrow 1$}}
\thicklines
\put(50,60){\circle{80}}
\put(50,20){\vector(0,-1){20}}
\put(47,100){\vector(-1,0){1}}
\put(10,57){\vector(0,-1){1}}
\put(30,25.5){\vector(2,-1){1}}
\put(70.5,25.5){\vector(2,1){1}}
\put(70,60){\vector(1,0){1}}
\put(33,80){\vector(1,0){1}}
\put(33,40){\vector(1,0){1}}
\qbezier(15,80)(15,80)(30,80)\qbezier(30,80)(40,80)(50,60)
\qbezier(15,40)(15,40)(30,40)\qbezier(30,40)(40,40)(50,60)
\qbezier(50,60)(50,60)(90,60)
\put(50,20){\lamnode}
\put(15,80){\lamnode}
\put(15,40){\lamnode}
\put(50,60){\appnode}
\put(90,60){\appnode}
\put(45,40){\makebox(0,0){\tiny$x$}}
\put(45,78){\makebox(0,0){\tiny$y$}}
\put(70,68){\makebox(0,0){\tiny$x\,y$}}
\put(92,40){\makebox(0,0){\tiny$f$}}
\put(100,92){\makebox(0,0){\tiny$f\,(x\,y)$}}
\put(-20,60){\makebox(0,0){\tiny$\lambda y.f\,(x\,y)$}}
\put(7,15){\makebox(0,0){\tiny$\lambda xy.f\,(x\,y)$}}
\put(88,3){\makebox(0,0){\tiny$\lambda fxy.f\,(x\,y)$}}
\end{picture}
\begin{picture}(50,100)
\end{picture}
\begin{picture}(100,120)
\put(50,115){\makebox(0,0){$\CC:2\rightarrow 2$}}
\thicklines
\put(50,60){\circle{80}}
\put(50,20){\vector(0,-1){20}}
\put(47,100){\vector(-1,0){1}}
\put(10,57){\vector(0,-1){1}}
\put(30,25.5){\vector(2,-1){1}}
\put(70.5,25.5){\vector(2,1){1}}
\put(90,63){\vector(0,1){1}}
\put(33,80){\vector(1,0){1}}
\put(33,40){\vector(1,0){1}}
\qbezier(15,80)(15,80)(30,80)\qbezier(30,80)(40,80)(50,60)\qbezier(50,60)(60,40)(70,40)\qbezier(70,40)(85,40)(85,40)
\qbezier(15,40)(15,40)(30,40)\qbezier(30,40)(40,40)(50,60)\qbezier(50,60)(60,80)(70,80)\qbezier(70,80)(85,80)(85,80)
\put(50,20){\lamnode}
\put(15,80){\lamnode}
\put(15,40){\lamnode}
\put(85,80){\appnode}
\put(85,40){\appnode}
\put(45,40){\makebox(0,0){\tiny$x$}}
\put(45,78){\makebox(0,0){\tiny$y$}}
\put(101,62){\makebox(0,0){\tiny$f\,y$}}
\put(88,28){\makebox(0,0){\tiny$f$}}
\put(95,95){\makebox(0,0){\tiny$f\,y\,x$}}
\put(-18,60){\makebox(0,0){\tiny$\lambda y.f\,y\,x$}}
\put(7,15){\makebox(0,0){\tiny$\lambda xy.f\,y\,x$}}
\put(86,3){\makebox(0,0){\tiny$\lambda fxy.f\,y\,x$}}
\end{picture}
\begin{picture}(50,120)
\end{picture}
\begin{picture}(100,120)
\put(50,115){\makebox(0,0){$\II:0\rightarrow 0$}}
\thicklines
\put(50,60){\circle{80}}
\put(50,20){\vector(0,-1){20}}
\put(47,100){\vector(-1,0){1}}
\put(50,20){\lamnode}
\put(98,55){\makebox(0,0){\tiny$f$}}
\put(70,3){\makebox(0,0){\tiny$\lambda f.f$}}
\end{picture}
\begin{picture}(40,120)
\end{picture}
\begin{picture}(100,120)
\put(50,115){\makebox(0,0){$P^\bullet:0\rightarrow1$}}
\thicklines
\put(50,60){\circle{80}}
\put(50,20){\vector(0,-1){20}}
\put(47,100){\vector(-1,0){1}}
\put(70,60){\vector(1,0){20}}
\put(50,60){\shade\color[rgb]{.7,.7,.7}\circle{40}}
\put(50,60){\makebox(0,0){$P$}}
\put(50,20){\lamnode}
\put(90,60){\appnode}
\put(98,85){\makebox(0,0){\tiny$f\,P$}}
\put(88,28){\makebox(0,0){\tiny$f$}}
\put(73,3){\makebox(0,0){\tiny$\lambda f.f\,P$}}
\end{picture}
\end{center}
They will be the basic primitives for the planar and linear combinatory algebras.

Now we describe a few simple constructions on terms with arity.
They will be of fundamental importance in describing the operad structure.

\paragraph{Adding lower strands}
For $M:m\rightarrow n$, we have $\BB\,M:m+1\rightarrow n+1$.
graphically, applying $\BB$ adds a new lower strand:

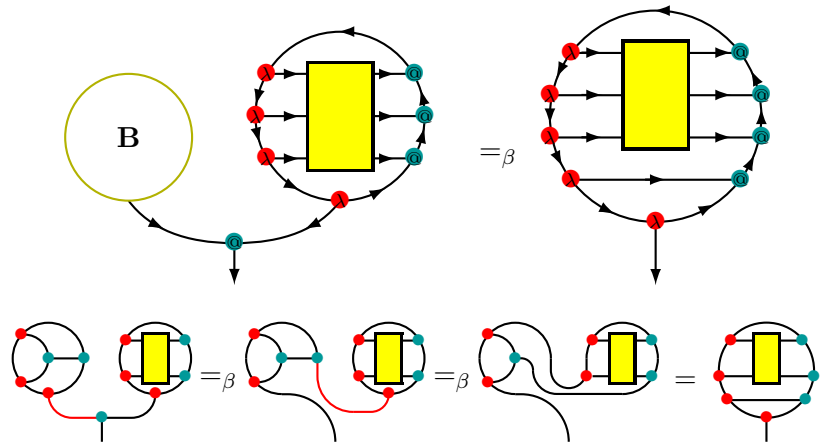
\begin{center}
\unitlength=.8pt
\begin{picture}(180,140)(0,0)
\thicklines
\qbezier(30,50)(45,30)(80,30)%
\qbezier(130,50)(115,30)(80,30)%

\put(80,30){\vector(0,-1){20}}
\put(30,80){\shade\color[rgb]{.7,.7,0}\circle{60}}%
\put(30,80){\makebox(0,0){$\BB$}}%
\put(80,50){\usebox{\boxp}}  
\rgbfbox{115}{65}{30}{50}{1}{1}{0}{}%

\put(80,30){\appnode}%
\put(46,37){\vector(2,-1){1}}
\put(114,37){\vector(-2,-1){1}}

\end{picture}
\begin{picture}(40,140)(0,0)
\put(20,70){\makebox(0,0){$=_\beta$}}
\end{picture}
\begin{picture}(100,140)(0,0)
\thicklines
\put(50,90){\circle{100}}
\rgbfbox{35}{75}{30}{50}{1}{1}{0}{}%

\qbezier(10,120)(35,120)(35,120) \qbezier(65,120)(65,120)(90,120)
\qbezier(0,100)(0,100)(35,100) \qbezier(65,100)(100,100)(100,100)
\qbezier(0,80)(0,80)(35,80) \qbezier(65,80)(100,80)(100,80)
{\keywordcolor \qbezier(10,60)(50,60)(90,60) }
\put(15,120){\vector(1,0){10}}
\put(15,100){\vector(1,0){10}}
\put(15,80){\vector(1,0){10}}

{\keywordcolor \put(45,60){\vector(1,0){10}} }

\put(85,120){\vector(1,0){1}}
\put(85,100){\vector(1,0){1}}
\put(85,80){\vector(1,0){1}}
\put(50,40){\vector(0,-1){30}}
\put(50,140){\vector(-1,0){1}}
\put(4,110){\vector(-1,-2){1}}
\put(0,87){\vector(0,-1){1}}
\put(5,68){\vector(1,-2){1}}

\put(96,110){\vector(-1,2){1}}
\put(100,93){\vector(0,1){1}}
\put(96,70){\vector(1,2){1}}
\put(30,44){\vector(2,-1){1}}
\put(77,48){\vector(2,1){1}}
\put(10,120){\lamnode}
\put(0,100){\lamnode}
\put(0,80){\lamnode}
\put(10,60){\lamnode}
\put(50,40){\lamnode}
\put(90,120){\appnode}
\put(100,100){\appnode}
\put(100,80){\appnode}
\put(90,60){\appnode}
\end{picture}
\end{center}
\begin{center}
\unitlength=.45pt
\begin{picture}(150,100)
\thicklines
\put(30,70){\oval(60,60)}
\put(120,70){\oval(60,60)}
\put(75,40){\oval(90,40)[br]} \put(75,40){\red\oval(90,40)[bl]} 
\put(75,0){\line(0,1){20}}
\put(7,70){\oval(46,40)[r]}
\put(30,70){\line(1,0){30}}
\put(7,90){\tinylamnode}
\put(7,50){\tinylamnode}
\put(30,40){\tinylamnode}
\put(30,70){\tinyappnode}
\put(60,70){\tinyappnode}
\put(75,20){\tinyappnode}
\rgbfbox{110}{50}{20}{40}{1}{1}{0}{}
\put(95,85){\line(1,0){15}}
\put(95,55){\line(1,0){15}}
\put(130,85){\line(1,0){15}}
\put(130,55){\line(1,0){15}}
\put(95,85){\tinylamnode}
\put(95,55){\tinylamnode}
\put(120,40){\tinylamnode}
\put(145,85){\tinyappnode}
\put(145,55){\tinyappnode}
\end{picture}
\begin{picture}(30,100)
\put(15,50){\makebox(0,0){$=_\beta$}}
\end{picture}
\begin{picture}(150,100)
\thicklines
\put(30,70){\oval(60,60)[t]}\put(30,70){\oval(60,60)[bl]}
\put(120,70){\oval(60,60)}
\put(7,70){\oval(46,40)[r]}
\put(30,70){\line(1,0){30}}
\put(30,0){\oval(90,80)[tr]}
\put(90,70){\red\oval(60,90)[bl]}
\put(90,40){\red\oval(60,30)[br]} 
\rgbfbox{110}{50}{20}{40}{1}{1}{0}{}
\put(95,85){\line(1,0){15}}
\put(95,55){\line(1,0){15}}
\put(130,85){\line(1,0){15}}
\put(130,55){\line(1,0){15}}
\put(7,90){\tinylamnode}
\put(7,50){\tinylamnode}
\put(30,70){\tinyappnode}
\put(60,70){\tinyappnode}
\put(95,85){\tinylamnode}
\put(95,55){\tinylamnode}
\put(145,85){\tinyappnode}
\put(145,55){\tinyappnode}
\put(120,40){\tinylamnode}
\end{picture}
\begin{picture}(30,100)
\put(15,50){\makebox(0,0){$=_\beta$}}
\end{picture}
\begin{picture}(150,100)
\thicklines
\put(30,70){\oval(60,60)[t]}\put(30,70){\oval(60,60)[bl]}
\put(120,70){\oval(60,60)[t]}\put(120,70){\oval(60,60)[br]}
\put(7,70){\oval(46,40)[r]}
\put(30,0){\oval(90,80)[tr]}
\put(75,70){\oval(30,50)[b]}
\put(30,60){\oval(30,20)[tr]}
\put(120,60){\oval(150,40)[bl]}
\rgbfbox{110}{50}{20}{40}{1}{1}{0}{}
\put(95,85){\line(1,0){15}}
\put(95,55){\line(1,0){15}}
\put(130,85){\line(1,0){15}}
\put(130,55){\line(1,0){15}}
\put(7,90){\tinylamnode}
\put(7,50){\tinylamnode}
\put(30,70){\tinyappnode}
\put(95,85){\tinylamnode}
\put(90,55){\tinylamnode}
\put(145,85){\tinyappnode}
\put(145,55){\tinyappnode}
\end{picture}
 %
\begin{picture}(35,100)
\put(15,50){\makebox(0,0){$=$}}
\end{picture}
\begin{picture}(80,100)
\thicklines
\put(40,60){\oval(80,80)}
\rgbfbox{30}{50}{20}{40}{1}{1}{0}{}
\put(10,85){\line(1,0){20}}
\put(0,55){\line(1,0){30}}
\put(50,85){\line(1,0){20}}
\put(50,55){\line(1,0){30}}
\put(10,35){\line(1,0){60}}
\put(40,0){\line(0,1){20}}
\put(10,85){\tinylamnode}
\put(0,55){\tinylamnode}
\put(7,35){\tinylamnode}
\put(40,20){\tinylamnode}
\put(70,85){\tinyappnode}
\put(80,55){\tinyappnode}
\put(73,35){\tinyappnode}
\end{picture}
\end{center}

\paragraph{Adding upper strands}
As we already noticed, $M:m\rightarrow n$ implies $M:m+1\rightarrow n+1$.
Graphically, it means that we can add upper strands for free:

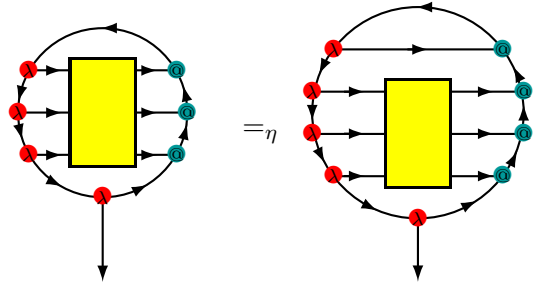
\begin{center}
\unitlength=.8pt
\begin{picture}(100,130)(80,10)
\thicklines
\put(130,50){\vector(0,-1){40}}
\put(80,50){\usebox{\boxp}}  
\rgbfbox{115}{65}{30}{50}{1}{1}{0}{}%
\end{picture}
\begin{picture}(40,130)(0,0)
\put(20,70){\makebox(0,0){$=_\eta$}}
\end{picture}
\begin{picture}(100,130)(0,10)
\thicklines
\put(50,90){\circle{100}}
\rgbfbox{35}{55}{30}{50}{1}{1}{0}{}%

\qbezier(10,60)(35,60)(35,60) \qbezier(65,60)(65,60)(90,60)
\qbezier(0,100)(0,100)(35,100) \qbezier(65,100)(100,100)(100,100)
\qbezier(0,80)(0,80)(35,80) \qbezier(65,80)(100,80)(100,80)
{\keywordcolor \qbezier(10,120)(50,120)(90,120) }
\put(15,60){\vector(1,0){10}}
\put(15,100){\vector(1,0){10}}
\put(15,80){\vector(1,0){10}}

{\keywordcolor \put(45,120){\vector(1,0){10}} }

\put(85,60){\vector(1,0){1}}
\put(85,100){\vector(1,0){1}}
\put(85,80){\vector(1,0){1}}
\put(50,40){\vector(0,-1){30}}
\put(50,140){\vector(-1,0){1}}
\put(4,110){\vector(-1,-2){1}}
\put(0,87){\vector(0,-1){1}}
\put(5,68){\vector(1,-2){1}}

\put(96,110){\vector(-1,2){1}}
\put(100,93){\vector(0,1){1}}
\put(96,70){\vector(1,2){1}}
\put(30,44){\vector(2,-1){1}}
\put(77,48){\vector(2,1){1}}
\put(10,120){\lamnode}
\put(0,100){\lamnode}
\put(0,80){\lamnode}
\put(10,60){\lamnode}
\put(50,40){\lamnode}
\put(90,120){\appnode}
\put(100,100){\appnode}
\put(100,80){\appnode}
\put(90,60){\appnode}
\end{picture}
\end{center}

\paragraph{Sequential composition}
As usual, let us write $\semcolor M\circ N$ for $\semcolor \BB\,M\,N=_{\beta}\lambda f.M\,(N\,f)$.  For $M:l\rightarrow m$ and $N:m\rightarrow n$, 
we have $M\circ N:l\rightarrow n$, the sequential composition of $M$ and $N$:

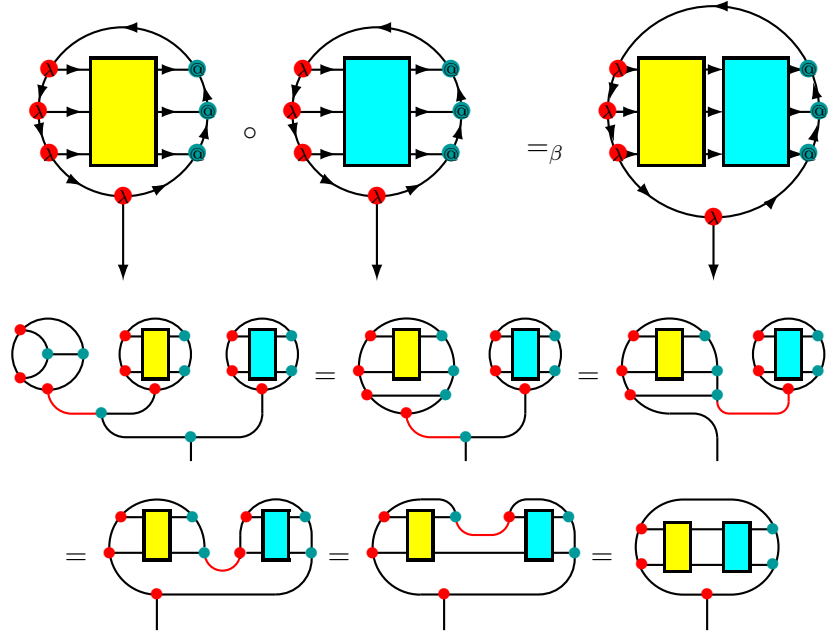
\begin{center}
\unitlength=.8pt
\begin{picture}(220,140)(0,0)
\thicklines

\put(50,50){\vector(0,-1){40}}
\put(170,50){\vector(0,-1){40}}
\put(0,50){\usebox{\boxp}}  
\rgbfbox{35}{65}{30}{50}{1}{1}{0}{}%
\put(120,50){\usebox{\boxp}} 
\rgbfbox{155}{65}{30}{50}{0}{1}{1}{}%
\put(110,80){\makebox(0,0){$\circ$}}

\end{picture}
\begin{picture}(50,140)(0,0)
\put(25,70){\makebox(0,0){$=_\beta$}}
\end{picture}
\begin{picture}(100,140)(0,0)
\thicklines
\put(50,90){\circle{100}}
\rgbfbox{15}{65}{30}{50}{1}{1}{0}{}%
\rgbfbox{55}{65}{30}{50}{0}{1}{1}{}%
\qbezier(5,110)(5,110)(15,110) \qbezier(85,110)(95,110)(95,110)
\qbezier(0,90)(0,90)(15,90) \qbezier(85,90)(100,90)(100,90)
\qbezier(5,70)(5,70)(15,70) \qbezier(85,70)(95,70)(95,70)
\put(5,110){\vector(1,0){10}}
\put(0,90){\vector(1,0){15}}
\put(5,70){\vector(1,0){10}}
\put(45,110){\vector(1,0){10}}
\put(45,90){\vector(1,0){10}}
\put(45,70){\vector(1,0){10}}
\put(92,110){\vector(1,0){1}}
\put(92,90){\vector(1,0){1}}
\put(92,70){\vector(1,0){1}}
\put(50,40){\vector(0,-1){30}}
\put(50,140){\vector(-1,0){1}}
\put(2,100){\vector(-1,-4){1}}
\put(2,80){\vector(1,-4){1}}
\put(98,100){\vector(-1,4){1}}
\put(98,80){\vector(1,4){1}}
\put(20,50){\vector(1,-1){1}}
\put(80,50){\vector(1,1){1}}
\put(5,110){\lamnode}
\put(0,90){\lamnode}
\put(5,70){\lamnode}
\put(50,40){\lamnode}
\put(95,110){\appnode}
\put(100,90){\appnode}
\put(95,70){\appnode}
\end{picture}
\end{center}

\begin{center}
\unitlength=.45pt
\begin{picture}(240,100)
\thicklines
\put(30,70){\oval(60,60)}
\put(120,70){\oval(60,60)}
\put(210,70){\oval(60,60)}
\put(75,40){\oval(90,40)[br]} \put(75,40){\red\oval(90,40)[bl]} 
\put(142.5,20){\oval(135,40)[b]} \put(210,20){\line(0,1){20}}
\put(150,0){\line(0,-1){20}}
\put(7,70){\oval(46,40)[r]}
\put(30,70){\line(1,0){30}}
\put(7,90){\tinylamnode}
\put(7,50){\tinylamnode}
\put(30,40){\tinylamnode}
\put(30,70){\tinyappnode}
\put(60,70){\tinyappnode}
\put(75,20){\tinyappnode}
\put(150,0){\tinyappnode}
\rgbfbox{110}{50}{20}{40}{1}{1}{0}{}
\put(95,85){\line(1,0){15}}
\put(95,55){\line(1,0){15}}
\put(130,85){\line(1,0){15}}
\put(130,55){\line(1,0){15}}
\rgbfbox{200}{50}{20}{40}{0}{1}{1}{}
\put(185,85){\line(1,0){15}}
\put(185,55){\line(1,0){15}}
\put(220,85){\line(1,0){15}}
\put(220,55){\line(1,0){15}}
\put(95,85){\tinylamnode}
\put(95,55){\tinylamnode}
\put(120,40){\tinylamnode}
\put(145,85){\tinyappnode}
\put(145,55){\tinyappnode}
\put(185,85){\tinylamnode}
\put(185,55){\tinylamnode}
\put(210,40){\tinylamnode}
\put(235,85){\tinyappnode}
\put(235,55){\tinyappnode}
\end{picture}
\begin{picture}(35,100)
\put(15,50){\makebox(0,0){$=$}}
\end{picture}
\begin{picture}(170,100)
\thicklines
\put(40,60){\oval(80,80)}
\rgbfbox{30}{50}{20}{40}{1}{1}{0}{}
\put(140,70){\oval(60,60)}
\put(90,20){\red\oval(100,40)[bl]} \put(90,20){\oval(100,40)[br]}
\put(140,20){\line(0,1){20}}
\put(90,0){\line(0,-1){20}}
\put(10,85){\line(1,0){20}}
\put(0,55){\line(1,0){30}}
\put(50,85){\line(1,0){20}}
\put(50,55){\line(1,0){30}}
\put(10,35){\line(1,0){60}}
\put(10,85){\tinylamnode}
\put(0,55){\tinylamnode}
\put(7,35){\tinylamnode}
\put(40,20){\tinylamnode}
\put(70,85){\tinyappnode}
\put(80,55){\tinyappnode}
\put(73,35){\tinyappnode}
\put(90,0){\tinyappnode}
\rgbfbox{130}{50}{20}{40}{0}{1}{1}{}
\put(115,85){\line(1,0){15}}
\put(115,55){\line(1,0){15}}
\put(150,85){\line(1,0){15}}
\put(150,55){\line(1,0){15}}
\put(115,85){\tinylamnode}
\put(115,55){\tinylamnode}
\put(140,40){\tinylamnode}
\put(165,85){\tinyappnode}
\put(165,55){\tinyappnode}
\end{picture}
\begin{picture}(35,100)
\put(15,50){\makebox(0,0){$=$}}
\end{picture}
\begin{picture}(170,100)
\thicklines
\put(40,60){\oval(80,80)[t]} \put(40,60){\oval(80,80)[bl]}
\rgbfbox{30}{50}{20}{40}{1}{1}{0}{}
\put(140,70){\oval(60,60)}
\put(110,30){\red\oval(60,20)[b]} \put(80,30){\line(0,1){25}}
\put(40,0){\oval(80,40)[tr]}
\put(140,30){\red\line(0,1){10}}
\put(80,0){\line(0,-1){20}}
\put(10,85){\line(1,0){20}}
\put(0,55){\line(1,0){30}}
\put(50,85){\line(1,0){20}}
\put(50,55){\line(1,0){30}}
\put(10,35){\line(1,0){70}}
\put(10,85){\tinylamnode}
\put(0,55){\tinylamnode}
\put(7,35){\tinylamnode}
\put(70,85){\tinyappnode}
\put(80,55){\tinyappnode}
\put(80,35){\tinyappnode}
\rgbfbox{130}{50}{20}{40}{0}{1}{1}{}
\put(115,85){\line(1,0){15}}
\put(115,55){\line(1,0){15}}
\put(150,85){\line(1,0){15}}
\put(150,55){\line(1,0){15}}
\put(115,85){\tinylamnode}
\put(115,55){\tinylamnode}
\put(140,40){\tinylamnode}
\put(165,85){\tinyappnode}
\put(165,55){\tinyappnode}
\end{picture}
\end{center}

\mbox{} 

\begin{center}
\unitlength=.45pt
\begin{picture}(35,100)
\put(15,50){\makebox(0,0){$=$}}
\end{picture}
\begin{picture}(170,100)
\thicklines
\put(40,60){\oval(80,80)[t]} \put(40,60){\oval(80,80)[bl]}
\rgbfbox{30}{50}{20}{40}{1}{1}{0}{}
\put(140,70){\oval(60,60)[t]}
\put(140,70){\oval(60,100)[br]}
\put(95,55){\red\oval(30,30)[b]} \put(110,50){\line(0,1){20}}
\put(40,20){\line(0,-1){30}}
\put(10,85){\line(1,0){20}}
\put(0,55){\line(1,0){30}}
\put(50,85){\line(1,0){20}}
\put(50,55){\line(1,0){30}}
\put(40,20){\line(1,0){100}}
\put(10,85){\tinylamnode}
\put(0,55){\tinylamnode}
\put(40,20){\tinylamnode}
\put(70,85){\tinyappnode}
\put(80,55){\tinyappnode}
\rgbfbox{130}{50}{20}{40}{0}{1}{1}{}
\put(115,85){\line(1,0){15}}
\put(115,55){\line(1,0){15}}
\put(150,85){\line(1,0){15}}
\put(150,55){\line(1,0){15}}
\put(115,85){\tinylamnode}
\put(110,55){\tinylamnode}
\put(165,85){\tinyappnode}
\put(170,55){\tinyappnode}
\end{picture}
\begin{picture}(35,100)
\put(15,50){\makebox(0,0){$=$}}
\end{picture}
\begin{picture}(170,100)
\thicklines
\put(40,60){\oval(80,80)[tl]} \put(40,60){\oval(80,80)[bl]}
\put(40,85){\oval(60,30)[tr]} 
\rgbfbox{30}{50}{20}{40}{1}{1}{0}{}
\put(140,70){\oval(60,60)[tr]} \put(140,85){\oval(50,30)[tl]}
\put(140,70){\oval(60,100)[br]}
\put(92.5,85){\red\oval(45,30)[b]}%
\put(60,20){\line(0,-1){30}}
\put(10,85){\line(1,0){20}}
\put(0,55){\line(1,0){30}}
\put(50,85){\line(1,0){20}}
\put(50,55){\line(1,0){80}}
\put(40,20){\line(1,0){100}}
\put(10,85){\tinylamnode}
\put(0,55){\tinylamnode}
\put(60,20){\tinylamnode}
\put(70,85){\tinyappnode}

\rgbfbox{130}{50}{20}{40}{0}{1}{1}{}
\put(115,85){\line(1,0){15}}

\put(150,85){\line(1,0){15}}
\put(150,55){\line(1,0){15}}
\put(115,85){\tinylamnode}

\put(165,85){\tinyappnode}
\put(170,55){\tinyappnode}
\end{picture}
\begin{picture}(35,100)
\put(15,50){\makebox(0,0){$=$}}
\end{picture}
\begin{picture}(120,100)
\thicklines
\put(60,60){\oval(120,80)}
\rgbfbox{25}{40}{20}{40}{1}{1}{0}{}
\put(60,20){\line(0,-1){30}}
\put(5,75){\line(1,0){20}}
\put(5,45){\line(1,0){20}}
\put(45,75){\line(1,0){30}}
\put(45,45){\line(1,0){30}}
\put(5,75){\tinylamnode}
\put(5,45){\tinylamnode}
\put(60,20){\tinylamnode}
\rgbfbox{75}{40}{20}{40}{0}{1}{1}{}
\put(95,75){\line(1,0){20}}
\put(95,45){\line(1,0){20}}
\put(115,75){\tinyappnode}
\put(115,45){\tinyappnode}
\end{picture}
\end{center}
The composition $\circ$ is associative, and 
$\II:n\rightarrow n$ serves as the unit.

\subsection{The Closed Operad Structure of $\ILam$}

Now we shall spell out the operad structure of $\ILam$.
For $\semcolor f_i\in\ILam(k_i)$ ($1\leq i\leq n$) and 
$\semcolor g\in\ILam(n)$, the composite
${\semcolor g(f_1,\dots,f_n)
\in\ILam(k_1+k_2+\dots + k_n)}$
is 
$\semcolor 
f_1\circ(\BB\,f_2)\circ\dots\circ(\BB^{n-1}\,f_n)\circ g
$ (Figure \ref{fig:composition}).
\begin{figure}

\begin{minipage}[b]{0.45\linewidth}
\begin{center}
\begin{picture}(150,120)(-10,-15)
\put(17,-5){\dashbox(26,100){}}
\put(47,-5){\dashbox(26,100){}}
\put(77,-5){\dashbox(26,100){}}
\put(107,-5){\dashbox(26,100){}}
\put(30,100){\makebox(0,0){\small$f_1$}}
\put(60,100){\makebox(0,0){\small$\BB\,f_2$}}
\put(90,100){\makebox(0,0){\small$\BB^2f_3$}}
\put(120,100){\makebox(0,0){\small$g$}}
\put(70,-15){\makebox(0,0){$f_1\circ (\BB\,f_2)\circ (\BB^2\,f_3)\circ g$}}
\thicklines
\rgbfbox{20}{0}{20}{30}{.7}{1}{.7}{$f_1$}%
\rgbfbox{50}{30}{20}{30}{1}{.7}{.7}{$f_2$}%
\rgbfbox{80}{60}{20}{30}{.7}{.7}{1}{$f_3$}%
\rgbfbox{110}{0}{20}{90}{1}{1}{0}{$g$}%
\multiput(10,65)(0,10){3}{\line(1,0){70}}
\multiput(10,35)(0,10){3}{\line(1,0){40}}
\multiput(10,5)(0,10){3}{\line(1,0){10}}
\put(40,15){\line(1,0){70}}
\put(70,45){\line(1,0){40}}
\put(100,75){\line(1,0){10}}
\put(130,45){\line(1,0){10}}
\put(3,72){$\bigg\{$}
\put(3,42){$\bigg\{$}
\put(3,12){$\bigg\{$}
\put(-7,72){\small$k_3$}
\put(-7,42){\small$k_2$}
\put(-7,12){\small$k_1$}
\end{picture}
\end{center}
\caption{Composition $g(f_1,f_2,f_3)$}\label{fig:composition}
\end{minipage}
\begin{minipage}[b]{0.45\linewidth}
\begin{center}
\unitlength=.5pt
\begin{picture}(140,180)
\put(30,175){\makebox(0,0){$\BB^m\,N$}}
\put(100,175){\makebox(0,0){$M$}}
\put(70,175){\makebox(0,0){$\circ$}}
\multiput(70,25)(0,5){27}{\makebox(0,0){$\cdot$}}
\thicklines
\put(70,90){\oval(120,140)}
%
\put(70,20){\vector(0,-1){20}}
\put(47,160){\vector(-1,0){1}}
\put(50,20){\vector(1,0){1}}
\rgbfbox{25}{95}{30}{40}{0}{1}{1}{}
\put(10,105){\line(1,0){15}}
\put(10,125){\line(1,0){15}}
\put(55,105){\line(1,0){75}}
\put(55,125){\line(1,0){75}}
\put(125,125){\vector(1,0){1}}
\put(125,105){\vector(1,0){1}}
\rgbfbox{85}{40}{30}{40}{1}{1}{0}{}
\put(10,70){\line(1,0){75}}
\put(10,50){\line(1,0){75}}
\put(115,70){\line(1,0){15}}
\put(115,50){\line(1,0){15}}
\put(65,70){\vector(1,0){1}}
\put(65,50){\vector(1,0){1}}
\put(125,70){\vector(1,0){1}}
\put(125,50){\vector(1,0){1}}
\put(10,125){\lamnode}
\put(10,105){\lamnode}
\put(130,125){\appnode}
\put(130,105){\appnode}
\put(70,20){\lamnode}
\put(10,70){\lamnode}
\put(10,50){\lamnode}
\put(130,70){\appnode}
\put(130,50){\appnode}
\put(-10,60){\makebox(0,0){\scriptsize$m$}}
\put(0,60){\makebox(0,0){$\Big\{$}}
\put(150,60){\makebox(0,0){\scriptsize$n$}}
\put(140,60){\makebox(0,0){$\Big\}$}}
\end{picture}
\begin{picture}(60,120)
\put(30,80){\makebox(0,0){$=$}}
\end{picture}
\begin{picture}(140,180)
\put(40,175){\makebox(0,0){$M$}}
\put(70,175){\makebox(0,0){$\circ$}}
\put(110,175){\makebox(0,0){$\BB^n\,N$}}
\multiput(70,25)(0,5){27}{\makebox(0,0){$\cdot$}}
\thicklines
\put(70,90){\oval(120,140)}
%
\put(70,20){\vector(0,-1){20}}
\put(47,160){\vector(-1,0){1}}
\put(50,20){\vector(1,0){1}}
\rgbfbox{85}{95}{30}{40}{0}{1}{1}{}
\put(10,105){\line(1,0){75}}
\put(10,125){\line(1,0){75}}
\put(115,105){\line(1,0){15}}
\put(115,125){\line(1,0){15}}
\put(65,125){\vector(1,0){1}}
\put(65,105){\vector(1,0){1}}
\put(127,125){\vector(1,0){1}}
\put(127,105){\vector(1,0){1}}
\put(127,70){\vector(1,0){1}}
\put(127,50){\vector(1,0){1}}
\rgbfbox{25}{40}{30}{40}{1}{1}{0}{}
\put(10,70){\line(1,0){15}}
\put(10,50){\line(1,0){15}}
\put(55,70){\line(1,0){75}}
\put(55,50){\line(1,0){75}}
\put(10,125){\lamnode}
\put(10,105){\lamnode}
\put(130,125){\appnode}
\put(130,105){\appnode}
\put(70,20){\lamnode}
\put(10,70){\lamnode}
\put(10,50){\lamnode}
\put(130,70){\appnode}
\put(130,50){\appnode}
\put(-10,60){\makebox(0,0){\scriptsize$m$}}
\put(0,60){\makebox(0,0){$\Big\{$}}
\put(150,60){\makebox(0,0){\scriptsize$n$}}
\put(140,60){\makebox(0,0){$\Big\}$}}
\end{picture}
\end{center}
\caption{The exchange law} \label{fig:exchange}
\end{minipage}

\end{figure}
With $\id=\II$, it is routine to see that this composition satisfies the
unit law and associativity of operads. 

Next, we look at the closed structure.
Let $\semcolor \app_{\IA}=\BB\in\ILam(2)$.
For $\semcolor t\in\ILam(m+1)$, let $\semcolor \lambda(t)\in\ILam(m)$ be $\semcolor (t\,\II)^\bullet\circ\BB^m$.
If $t$ is $\lambda fx_1\dots  x_m x_{m+1}.f\,M$,  $\lambda(t)$ is
$\lambda fx_1\dots  x_m.f\,(\lambda x_{m+1}.M)$.
Then $\semcolor \lambda(t)$ is the unique element satisfying 
${\semcolor t=\app_{\ILam}(\lambda(t),\id_{\ILam})=\lambda(t)\circ\BB.}$
Hence we conclude that $\ILam$ is a closed operad.

\subsection{Towards Internal Operads of Combinatory Algebras}

We have seen that, in the case of the $\lambda$-calculus and 
its graphical presentation, the following constructs
are essential in defining the internal operad $\ILam$:
the basic operators
$$\semcolor 
\infer{\BB:2\rightarrow 1}{}
~~~~~~~
\infer{\II:0\rightarrow 0}{}
~~~~~~~
\infer{P^\bullet:0\rightarrow 1}{P~\mbox{a~closed~term}}
$$%
and the composition as well as adding strands
$$\semcolor 
\infer{M\circ N: l\rightarrow n}
          {M:l\rightarrow m & N:m\rightarrow n}
~~~~~~
\infer{\BB\,M:m+1\rightarrow n+1}{M:m\rightarrow n}
~~~~~~
\infer{M:m+1\rightarrow n+1}{M:m\rightarrow n}
$$%
So far,
terms with arity are defined using head normal forms. 
However, it is possible to characterize them
just by using equations involving 
$\semcolor \BB$, $\semcolor \II$, $\semcolor (\_)^\bullet$,
with no mention to head normal forms 
as follows.
\begin{proposition}\label{prop:arity-m-1}
A closed $\lambda$-term $M$ is of arity $m\rightarrow 1$ (or $M\in \ILam(m)$)
iff
$(M\,\II)^\bullet\circ\BB^m=_{\beta\eta}M$
iff 
$M^\bullet\circ\BB^{m+1}=_{\beta\eta}(\BB\,M)\circ\BB$.
\end{proposition}
\noindent
Indeed, for $M=\lambda fx_1\dots x_m.f\,N$, it is not hard to
verify 
$M^\bullet\circ\BB^{m+1}=(\BB\,M)\circ\BB=
\lambda fgx_1\dots x_m.f\,(g\,N)$.
$M^\bullet\circ\BB^{m+1}=(\BB\,M)\circ\BB$ implies
$((M\,\II)^\bullet\circ\BB^m)\,f=(M^\bullet\circ\BB^{m+1})\,f\,\II=((\BB\,M)\circ\BB)\,f\,\II=M\,f$, hence $(M\,\II)^\bullet\circ\BB^m=M$.
Finally, $(M\,\II)^\bullet\circ\BB^m=M$ implies
$M\,f\,x_1\,\dots\,x_m=f\,(M\,\II\,x_1\,\dots\,x_m)$, hence 
$M=\lambda fx_1\dots x_m.f\,(M\,\II\,x_1\,\dots\,x_m)$.
More generally,  we have
\begin{proposition}\label{prop:arity-m-n}
A closed $\lambda$-term $M$ is of arity $m\rightarrow n$ 
iff 
$M^\bullet\circ\BB^{m+1}=_{\beta\eta}(\BB\,M)\circ\BB^n$.
\end{proposition}
These suggest that the internal operad  construction can be carried out 
in any applicative structure with 
$\semcolor \BB$, $\semcolor \II$ and $\semcolor (\_)^\bullet$
which validates the $\beta\eta$-equality (hence combinatory complete and
extensional). 

We conclude this section by noting that the condition
$\semcolor M^\bullet\circ\BB^{m+1}=_{\beta\eta}(\BB\,M)\circ\BB^n$
of Proposition \ref{prop:arity-m-n}
can be understood as an {\kc exchange law}
${\semcolor 
(\BB^m N)\circ M= M\circ (\BB^n N)}
$
as depicted in Figure \ref{fig:exchange}.

\section{Planar Combinatory Algebras}
\label{sec:planar}

\subsection{The Operad of Planar Polynomials}

Given an applicative structure $\semcolor \calA$, we construct an operad
$\semcolor \CA$, where $\semcolor \CA(m)$ is the smallest
class of functions from $\semcolor \mathcal{A}^m$ to $\semcolor \mathcal{A}$
such that
\begin{itemize}
\item $\semcolor \CA(0)=\mathcal{A}$,
\item $\semcolor \id_\mathcal{A}\in\CA(1)$, and
\item $\semcolor t_1\cdot t_2\in \CA(m+n)$
 for $\semcolor t_1\in \CA(m)$ and
 $\semcolor t_2\in\CA(n)$, where\\
 $\semcolor (t_1\cdot t_2)(x_1,\dots,x_m,y_1,\dots,y_n)=t_1(x_1,\dots,x_m)\cdot t_2(y_1,\dots,y_n).$%
\end{itemize}%
The elements of $\semcolor \CA(m)$ are {\em\keywordcolor planar polynomials} with $m$ variables. 
(If we allow pre-composing permutations, we have
{\kc linear polynomials}. 
If projections and duplications are allowed, we have
the usual (non-linear) polynomials.)

The identity function $\semcolor \id$ represents an occurrence of a variable.
$\semcolor \mathbf{app}=\id\cdot\id\in\CA(2)$ corresponds to the application:
$\semcolor \app(p,q)=p\cdot q$.
Two planar polynomials with $m$-variables are equal when
they are equal as 
functions from $\semcolor \mathcal{A}^m$ to $\semcolor \mathcal{A}$.

\subsection{$\BIdot$-algebras as Planarly Combinatory Complete Applicative Structures}

Suppose that $\semcolor \calA$ is an applicative structure which is {\kc combinatory complete}
with respect to the planar polynomials $\semcolor \CA$.
That is, for any $\semcolor p\in\PA(n+1)$, there exists $\semcolor \lambda^*(p)\in\PA(n)$
such that $\semcolor \lambda^*(p)\cdot\id=p$.
In $\semcolor \calA$, we have 
\begin{itemize}
\item $\semcolor \II=\lambda^*(\id)\in\calA$ which satisfies $\II\cdot a=a$
\item $\semcolor \BB=\lambda^*(\lambda^*(\lambda^*(\mathbf{app}(\id,\mathbf{app}))))\in\calA$ satisfying $\BB\cdot a\cdot b \cdot c=a\cdot(b\cdot c)$
\item $\semcolor a^\bullet=\lambda^*(\mathbf{app}(\id,a))\in\calA$ for $\semcolor a\in\calA$,
which satisfies $a^\bullet\cdot b=b\cdot a$
\end{itemize}

Conversely, if an applicative structure $\semcolor \calA$ has elements $\semcolor \II$, $\semcolor \BB$ and $\semcolor a^\bullet$ for all $\semcolor a\in\calA$ satisfying 
$\II\,a=a$, $\BB\,a\,b\,c=a\,(b\,c)$ and $a^\bullet\,b=b\,a$,
$\semcolor \calA$ is combinatory complete with respect to the planar polynomials:
$$\semcolor 
\begin{array}{rclrcl}
\lambda^*(\id) &=& \II
~~~~~~~
&
\lambda^*(\app(t_1,t_2)) &=&
\left\{
\begin{array}{ll}
\app(\app(\BB,t_2^\bullet),\lambda^*(t_1)) & t_2\in\calA\\
\app(\app(\BB,t_1),\lambda^*(t_2)) & \mbox{otherwise}
\end{array}
\right.
\end{array}
$$%
Following Tomita \cite{Tom21}, we call such an $\semcolor \calA$ a {\em $\BIdot$-algebra}. Thus, $\BIdot$-algebras are precisely the
planarly combinatory complete applicative structures.
There are several interesting $\BIdot$-algebras  including: 
the term model of the planar $\lambda$-calculus 
modulo
$\beta$- or $\beta\eta$-equality;
reflexive objects in monoidal closed categories; and 
models of Moggi's computational $\lambda$-calculus.
Originally, $\BIdot$-algebras were introduced in 
Tomita's study on {\kc non-symmetric (or planar) realizability}.
One of the central results in that context is
that the  assemblies on a
$\BIdot$-algebra form a closed multicategory. 
See \cite{Tom21,Tom22} for further details, variations and examples.

\subsection{Extensional $\BIdot$-algebras}

Planar combinatory completeness implies  a natural 
interpretation $\sem{\_}$ of the planar $\lambda$-calculus in a $\BIdot$-algebra, which 
validates the $\beta$-rule: $\sem{(\lambda x.M)\,N}=\sem{M[x:=N]}$.
However, the translation is in general {\kc not} sound for the {\kc $\eta$-equality}:
$\sem{\lambda x.M\,x}~\equiv~\lambda^* x.\sem{M\,x}\equiv\BB\,\sem{M}\,\II$, which may not agree with $\sem{M}$. Also the $\xi$-rule does not hold:
$\sem{M}=\sem{N}$ does not imply $\sem{\lambda x.M}=\sem{\lambda x.N}$ in general.
To remedy this, we introduce additional axioms to $\BIdot$-algebras:
\begin{definition}
A $\BIdot$-algebra is {\em  extensional} when it satisfies the following axioms.
$$
\begin{array}{|rcll|}
\hline
\BB\,\II &=& \II &\mathrm{(BI)}\\
(a\,b)^\bullet &=& \BB\,b^\bullet\,(\BB\,a^\bullet\,\BB)&\mathrm{(app\bullet)}\\
\BB\,\BB^\bullet\,(\BB\,\BB\,(\BB\,\BB\,\BB))
&=&
\BB\,(\BB\,\BB)\,\BB & \mathrm{(B\bullet)}\\
\BB\,\II^\bullet\,\BB &=& \II  & \mathrm{(I\bullet)}\\
\BB\,a^{\bullet\bullet}\,\BB &=&\BB\,(\BB\,a^\bullet)\,\BB & \mathrm{(\bullet\bullet)}\\
\hline
\end{array}
$$ 
\end{definition}
\noindent
Extensionality implies a lot.

\begin{lemma}
In an extensional $\BIdot$-algebra, 
the composition $\semcolor a\circ b=\BB\,a\,b$ is associative, and 
$\II$ is its unit: that is,
$a\circ(b\circ c)=(a\circ b)\circ c$ and $\II\circ a = a = a\circ\II$ hold.
\end{lemma}

\comment{
Firstly, in an extensional $\BIdot$-algebra, 
the composition $\semcolor a\circ b=\BB\,a\,b$ is associative, and 
$\II$ is its unit: that is,
$a\circ(b\circ c)=(a\circ b)\circ c$ and $\II\circ a = a = a\circ\II$ hold.
}

The extensional equality is a congruence for the $\lambda^*$-abstraction, and it follows that  soundness for the  $\beta\eta$-equality holds:
$M=_{\beta\eta} N$ in the planar $\lambda$-calculus  implies $\sem{M}=\sem{N}$
in any extensional $\BIdot$-algebra.
Also, it is routine to see that the closed term model of the planar $\lambda\beta\eta$-calculus is an
extensional $\BIdot$-algebra.
So are the term models of the $\lambda\beta\eta$-calculus,
linear $\lambda\beta\eta$-calculus, and even the braided $\lambda\beta\eta$-calculus. As a result, completeness for the $\beta\eta$-equality holds:

\begin{proposition}
$M=_{\beta\eta} N$ in the planar $\lambda$-calculus if and only if $\sem{M}=\sem{N}$ for all extensional $\BIdot$-algebras.
\end{proposition}

Moreover, any extensional $\BIdot$-algebra has 
an internally defined isomorphic $\BIdot$-algebra in itself:

\begin{proposition}
For an extensional $\BIdot$-algebra $\calA$, 
$\semcolor \mathcal{A}^\bullet\equiv\{a^\bullet~|~a\in\mathcal{A}\}$ is
a $\BIdot$-algebra with $a\cdot_{\calA^\bullet} b=b\circ a \circ\BB$,
$\BB_{\calA^\bullet}=\BB^\bullet$, $\II_{\calA^\bullet}=\II^\bullet$ and
$a^{\bullet_{\calA^\bullet}}=a^\bullet$, which is 
 isomorphic to  $\semcolor \mathcal{A}$ via 
$a\mapsto a^\bullet:\calA\stackrel{\cong}{\longrightarrow}\calA^\bullet$
and 
$b\mapsto b\,\II:\calA^\bullet\stackrel{\cong}{\longrightarrow}\calA.$
\end{proposition}
Indeed, the axiom $\mathrm{(app\bullet)}$ states that
$(a\cdot b)^\bullet= a^\bullet\cdot_{\calA^\bullet} b^\bullet$ holds,
and 
$\mathrm{(B\bullet)}$, $\mathrm{(I\bullet)}$ and $\mathrm{(\bullet\bullet)}$ imply that axioms for $\BB$, $\II$ and $(-)^\bullet$ hold in $\calA^\bullet$.
If we follow the graphical presentation in the previous section, the last three axioms can be depicted as follows, which might be more understandable:

$$
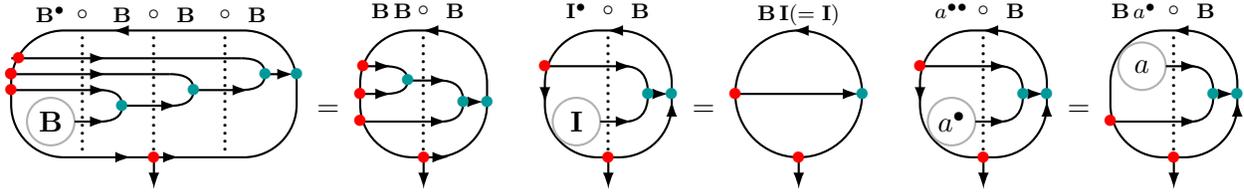
\unitlength=.6pt
\begin{picture}(200,120)(-10,0)
\thicklines
\put(100,60){\oval(180,80)}

\put(35,42.5){\shade\color[rgb]{.7,.7,.7}\circle{30}}\put(35,42.5){\makebox(0,0){$\BB$}}

\qbezier(65,62.5)(80,62.5)(80,52.5)
\qbezier(65,42.5)(80,42.5)(80,52.5)
\qbezier(110,72.5)(125,72.5)(125,62.5)
\qbezier(110,52.5)(125,52.5)(125,62.5)
\qbezier(155,82.5)(170,82.5)(170,72.5)
\qbezier(155,62.5)(170,62.5)(170,72.5)
\put(10,82.5){\vector(1,0){60}} \put(65,82.5){\line(1,0){90}}
\put(10,72.5){\vector(1,0){60}} \put(65,72.5){\line(1,0){45}}
\put(10,62.5){\vector(1,0){60}}
\put(50,42,5){\vector(1,0){20}}
\put(80,52.5){\vector(1,0){35}}
\put(125,62.5){\vector(1,0){35}}
\put(170,72.5){\vector(1,0){18}}
\put(100,20){\vector(0,-1){20}}
\put(85,20){\vector(1,0){1}}
\put(115,20){\vector(1,0){1}}
\put(75,100){\vector(-1,0){1}}
\put(100,20){\tinylamnode}
\put(10,72.5){\tinylamnode}
\put(80,52.5){\tinyappnode}
\put(125,62.5){\tinyappnode}
\put(10,62.5){\tinylamnode}
\put(10,72.5){\tinylamnode}
\put(15,82.5){\tinylamnode}
\put(170,72.5){\tinyappnode}
\put(190,72.5){\tinyappnode}
\multiput(55,25)(0,5){15}{\makebox(0,0){$\cdot$}}
\multiput(100,25)(0,5){15}{\makebox(0,0){$\cdot$}}
\multiput(145,25)(0,5){15}{\makebox(0,0){$\cdot$}}
\put(35,110){\makebox(0,0){\scriptsize$\BB^\bullet$}}
\put(80,110){\makebox(0,0){\scriptsize$\BB$}}
\put(120,110){\makebox(0,0){\scriptsize$\BB$}}
\put(165,110){\makebox(0,0){\scriptsize$\BB$}}
\put(55,110){\makebox(0,0){\scriptsize$\circ$}}
\put(100,110){\makebox(0,0){\scriptsize$\circ$}}
\put(145,110){\makebox(0,0){\scriptsize$\circ$}}
\end{picture}
\unitlength=.6pt
\begin{picture}(40,100)
\put(20,50){\makebox(0,0){$=$}}
\end{picture}
\unitlength=.6pt
\begin{picture}(80,120)(10,0)
\thicklines
\put(50,60){\oval(80,80)}

\qbezier(60,68.5)(75,68.5)(75,55)
\qbezier(60,42.5)(75,42.5)(75,55)
\qbezier(25,77.5)(40,77.5)(40,68.5)
\qbezier(25,60)(40,60)(40,68.5)
\put(75,55){\vector(1,0){13}}
\put(50,20){\vector(0,-1){20}}
\put(10,42.5){\vector(1,0){55}}
\put(40,68.5){\vector(1,0){25}}
\put(10,60){\vector(1,0){20}}
\put(10,77.5){\vector(1,0){20}}
\put(50,100){\vector(-1,0){1}}
\put(65,20){\vector(1,0){1}}
\put(50,20){\tinylamnode}
\put(12,77.5){\tinylamnode}
\put(10,60){\tinylamnode}
\put(10,42.5){\tinylamnode}
\put(40,68.5){\tinyappnode}
\put(75,55){\tinyappnode}
\put(90,55){\tinyappnode}
\multiput(50,25)(0,5){15}{\makebox(0,0){$\cdot$}}
\put(30,112.5){\makebox(0,0){\scriptsize$\BB\,\BB$}}
\put(70,112.5){\makebox(0,0){\scriptsize$\BB$}}
\put(50,112.5){\makebox(0,0){\scriptsize$\circ$}}
\end{picture}
~~~~~~
\unitlength=.6pt
\begin{picture}(80,115)(10,0)
\thicklines
\put(50,60){\oval(80,80)}
\put(30,42.5){\shade\color[rgb]{.7,.7,.7}\circle{30}}\put(30,42.5){\makebox(0,0){$\II$}}
\qbezier(60,77.5)(75,77.5)(75,60)
\qbezier(60,42.5)(75,42.5)(75,60)
\put(75,60){\vector(1,0){13}}
\put(50,20){\vector(0,-1){20}}
\put(10,77.5){\vector(1,0){55}}
\put(45,42.5){\vector(1,0){20}}
\put(50,100){\vector(-1,0){1}}
\put(90,55){\vector(0,1){1}}
\put(10,55){\vector(0,-1){1}}
\put(50,20){\tinylamnode}
\put(10,77.5){\tinylamnode}
\put(75,60){\tinyappnode}
\put(90,60){\tinyappnode}
\multiput(50,25)(0,5){15}{\makebox(0,0){$\cdot$}}
\put(30,112.5){\makebox(0,0){\scriptsize$\II^\bullet$}}
\put(70,112.5){\makebox(0,0){\scriptsize$\BB$}}
\put(50,112.5){\makebox(0,0){\scriptsize$\circ$}}
\end{picture}
\unitlength=.6pt
\begin{picture}(40,100)
\put(20,50){\makebox(0,0){$=$}}
\end{picture}
\unitlength=.6pt
\begin{picture}(80,115)(10,0)
\thicklines
\put(50,60){\circle{80}}
\put(50,20){\vector(0,-1){20}}
\put(47,100){\vector(-1,0){1}}
\put(10,60){\vector(1,0){78}}
\put(10,60){\tinylamnode}
\put(90,60){\tinyappnode}
\put(50,20){\tinylamnode}
\put(50,110){\makebox(0,0){\scriptsize$\BB\,\II(=\II)$}}
\end{picture}
~~~~~~
\unitlength=.6pt
\begin{picture}(80,115)(10,0)
\thicklines
\put(50,60){\oval(80,80)}
\put(30,42.5){\shade\color[rgb]{.7,.7,.7}\circle{30}}\put(30,42.5){\makebox(0,0){$a^\bullet$}}
\qbezier(60,77.5)(75,77.5)(75,60)
\qbezier(60,42.5)(75,42.5)(75,60)
\put(75,60){\vector(1,0){13}}
\put(50,20){\vector(0,-1){20}}
\put(10,77.5){\vector(1,0){55}}
\put(45,42.5){\vector(1,0){20}}
\put(50,100){\vector(-1,0){1}}
\put(90,55){\vector(0,1){1}}
\put(10,55){\vector(0,-1){1}}
\put(50,20){\tinylamnode}
\put(10,77.5){\tinylamnode}
\put(75,60){\tinyappnode}
\put(90,60){\tinyappnode}
\multiput(50,25)(0,5){15}{\makebox(0,0){$\cdot$}}
\put(30,112.5){\makebox(0,0){\scriptsize$a^{\bullet\bullet}$}}
\put(70,112.5){\makebox(0,0){\scriptsize$\BB$}}
\put(50,112.5){\makebox(0,0){\scriptsize$\circ$}}
\end{picture}
\unitlength=.6pt
\begin{picture}(40,100)
\put(20,50){\makebox(0,0){$=$}}
\end{picture}
\unitlength=.6pt
\begin{picture}(80,115)(10,0)
\thicklines
\put(50,60){\oval(80,80)}
\put(30,77.5){\shade\color[rgb]{.7,.7,.7}\circle{30}}\put(30,77.5){\makebox(0,0){$a$}}
\qbezier(60,77.5)(75,77.5)(75,60)
\qbezier(60,42.5)(75,42.5)(75,60)
\put(75,60){\vector(1,0){13}}
\put(50,20){\vector(0,-1){20}}
\put(10,42.5){\vector(1,0){55}}
\put(45,77.5){\vector(1,0){20}}
\put(50,100){\vector(-1,0){1}}
\put(90,55){\vector(0,1){1}}
\put(50,20){\tinylamnode}
\put(10,42.5){\tinylamnode}
\put(75,60){\tinyappnode}
\put(90,60){\tinyappnode}
\multiput(50,25)(0,5){15}{\makebox(0,0){$\cdot$}}
\put(25,112.5){\makebox(0,0){\scriptsize$\BB\,a^{\bullet}$}}
\put(70,112.5){\makebox(0,0){\scriptsize$\BB$}}
\put(50,112.5){\makebox(0,0){\scriptsize$\circ$}}
\end{picture}
$$
Actually, we have chosen these axioms following the graphical intuition.
The internally defined $\BIdot$-algebra $\calA^\bullet$ is part of the structure of the internal operad to be spelled out below.

\section{Internal Operads}
\label{sec:internal}

As explained in Section \ref{sec:motivation},
the idea of internal operads of a combinatory algebra $\calA$
was to use elements of arity $m\rightarrow1$
(Figure \ref{fig:arity-m-1}) 
\begin{figure}

\begin{minipage}[b]{0.45\linewidth}
\begin{center}
\unitlength=.6pt
\begin{picture}(100,120)(0,20)
\thicklines
\put(50,90){\circle{100}}
\put(20,65){\framebox(60,50){}}

\qbezier(5,110)(5,110)(20,110) 
\qbezier(0,90)(0,90)(20,90) \qbezier(80,90)(100,90)(100,90)
\qbezier(5,70)(5,70)(20,70) 
\put(5,110){\vector(1,0){15}}
\put(0,90){\vector(1,0){20}}
\put(5,70){\vector(1,0){15}}

\put(92,90){\vector(1,0){1}}
\put(50,40){\vector(0,-1){20}}
\put(50,140){\vector(-1,0){1}}
\put(2,100){\vector(-1,-4){1}}
\put(2,80){\vector(1,-4){1}}
\put(98,100){\vector(-1,4){1}}
\put(98,80){\vector(1,4){1}}
\put(20,50){\vector(1,-1){1}}
\put(80,50){\vector(1,1){1}}
\put(5,110){\lamnode}
\put(0,90){\lamnode}
\put(5,70){\lamnode}
\put(50,40){\lamnode}
\put(100,90){\appnode}
%
\put(-10,90){\makebox(0,0){$\Bigg\{$}}
\put(-34,90){\makebox(0,0){$m$~~~s}}
\put(-27,91){\lamnode}
\end{picture}
\end{center}
\caption{Elements of arity $m\rightarrow 1$}\label{fig:arity-m-1}
\end{minipage}
\begin{minipage}[b]{0.45\linewidth}
\begin{center}
\unitlength=.6pt
\begin{picture}(100,120)(0,20)
\thicklines
\put(50,90){\oval(200,100)}

\put(-10,60){\shade\color[rgb]{.7,.7,.7}\circle{30}}\put(-10,60){\makebox(0,0){$a$}}
\qbezier(5,60)(30,60)(30,70)
\put(50,40){\vector(0,-1){20}}
\put(50,140){\vector(-1,0){1}}
\qbezier(-50,110)(-50,110)(80,110)\qbezier(80,110)(110,110)(110,90)
\qbezier(-50,90)(-50,90)(40,90)\qbezier(40,90)(70,90)(70,80)
\qbezier(-50,70)(-35,70)(-30,75)\qbezier(-30,75)(-25,80)(-10,80)
\qbezier(-10,80)(-10,80)(10,80)\qbezier(10,80)(20,80)(30,70)
\qbezier(110,90)(110,90)(150,90)
\qbezier(70,80)(110,80)(110,90)
\qbezier(30,70)(70,70)(70,80)
\put(-50,110){\lamnode}
\put(-50,90){\lamnode}
\put(-50,70){\lamnode}
\put(50,40){\lamnode}
\put(150,90){\appnode}
\put(110,90){\appnode}
\put(70,80){\appnode}
\put(30,70){\appnode}
\put(-60,90){\makebox(0,0){$\Bigg\{$}}
\put(-84,90){\makebox(0,0){$m$~~~s}}
\put(-77,91){\lamnode}
\end{picture}
\end{center}
\caption{$a^\bullet\circ\BB^m:m\rightarrow 1$} \label{fig:a-bullet-B-m}
\end{minipage}

\end{figure}
as polynomials with $m$ variables. Thanks to combinatory completeness,
such elements of arity $m\rightarrow 1$ are equal to elements of the form
\comment{
\begin{center}
\unitlength=.6pt
\begin{picture}(100,120)(0,20)
\thicklines
\put(50,90){\oval(200,100)}

\put(-10,60){\shade\color[rgb]{.7,.7,.7}\circle{30}}\put(-10,60){\makebox(0,0){$a$}}
\qbezier(5,60)(30,60)(30,70)
\put(50,40){\vector(0,-1){20}}
\put(50,140){\vector(-1,0){1}}
\qbezier(-50,110)(-50,110)(80,110)\qbezier(80,110)(110,110)(110,90)
\qbezier(-50,90)(-50,90)(40,90)\qbezier(40,90)(70,90)(70,80)
\qbezier(-50,70)(-35,70)(-30,75)\qbezier(-30,75)(-25,80)(-10,80)
\qbezier(-10,80)(-10,80)(10,80)\qbezier(10,80)(20,80)(30,70)
\qbezier(110,90)(110,90)(150,90)
\qbezier(70,80)(110,80)(110,90)
\qbezier(30,70)(70,70)(70,80)
\put(-50,110){\lamnode}
\put(-50,90){\lamnode}
\put(-50,70){\lamnode}
\put(50,40){\lamnode}
\put(150,90){\appnode}
\put(110,90){\appnode}
\put(70,80){\appnode}
\put(30,70){\appnode}
\put(-60,90){\makebox(0,0){$\Bigg\{$}}
\put(-84,90){\makebox(0,0){$m$~~~s}}
\put(-77,91){\lamnode}
\end{picture}
\end{center}
}
$\semcolor  a^\bullet\circ\BB^m$ for some $\semcolor a\in\calA$
(Figure \ref{fig:a-bullet-B-m}).

\subsection{Internal Operads of Extensional $\BIdot$-algebras}

For an extensional $\BIdot$-algebra $\semcolor \calA$,
we define a closed operad $\semcolor \IA$, which we shall call the {\em\keywordcolor  internal operad} of $\semcolor \mathcal{A}$, by
$\IA(m)=\{a^\bullet\circ\BB^m~|~a\in\mathcal{A}\}
$
with 
$\semcolor \id_{\IA}=\II=\II^\bullet\circ\BB\in\IA(1)$ and 
$\semcolor \app_{\IA}=\BB=\II^\bullet\circ\BB\circ\BB\in\IA(2)$.
For $\semcolor f_i\in\IA(k_i)$ ($1\leq i\leq n$) and 
$\semcolor g\in\IA(n)$, the composite
${\semcolor g(f_1,\dots,f_n)
\in\IA(k_1+k_2+\dots + k_n)}$
is 
$\semcolor 
f_1\circ(\BB\,f_2)\circ\dots\circ(\BB^{n-1}\,f_n)\circ g.
$
\comment{
\begin{center}
\begin{picture}(150,120)(-10,-15)
\put(17,-5){\dashbox(26,100){}}
\put(47,-5){\dashbox(26,100){}}
\put(77,-5){\dashbox(26,100){}}
\put(107,-5){\dashbox(26,100){}}
\put(30,100){\makebox(0,0){\small$f_1$}}
\put(60,100){\makebox(0,0){\small$\BB\,f_2$}}
\put(90,100){\makebox(0,0){\small$\BB^2f_3$}}
\put(120,100){\makebox(0,0){\small$g$}}
\put(70,-15){\makebox(0,0){$f_1\circ (\BB\,f_2)\circ (\BB^2\,f_3)\circ g$}}
\thicklines
\rgbfbox{20}{0}{20}{30}{.7}{1}{.7}{$f_1$}%
\rgbfbox{50}{30}{20}{30}{1}{.7}{.7}{$f_2$}%
\rgbfbox{80}{60}{20}{30}{.7}{.7}{1}{$f_3$}%
\rgbfbox{110}{0}{20}{90}{1}{1}{0}{$g$}%
\multiput(10,65)(0,10){3}{\line(1,0){70}}
\multiput(10,35)(0,10){3}{\line(1,0){40}}
\multiput(10,5)(0,10){3}{\line(1,0){10}}
\put(40,15){\line(1,0){70}}
\put(70,45){\line(1,0){40}}
\put(100,75){\line(1,0){10}}
\put(130,45){\line(1,0){10}}
\put(3,72){$\bigg\{$}
\put(3,42){$\bigg\{$}
\put(3,12){$\bigg\{$}
\put(-7,72){\small$k_3$}
\put(-7,42){\small$k_2$}
\put(-7,12){\small$k_1$}
\end{picture}
\end{center}
}
For closedness, 
for $\semcolor t\in\IA(m+1)$, let $\semcolor \lambda(t)\in\IA(m)$ be $\semcolor (t\,I)^\bullet\circ\BB^m$.
$\semcolor \lambda(t)$ is the unique element satisfying 
${\semcolor t=\app_{\IA}(\lambda(t),\id_{\IA})=\lambda(t)\circ\BB.}$
(For verifying the closedness, it is useful to notice that $a\in\IA(m)$ if and only if 
$a=(a\,\II)^\bullet\circ\BB^m$ holds --- indeed, for $x=a^\bullet\circ\BB^m$,
$x\,\II=a^\bullet\,(\BB^m\,\II)=a^\bullet\,\II=a$, hence $(x\,\II)^\bullet\circ\BB^m=x$.) 

\begin{proposition}
For any extensional $\BIdot$-algebra $\semcolor \calA$,  
$\semcolor \IA$ is a closed operad s.t. $\semcolor \IA(0)=\calA^\bullet\cong \calA$.
That is, $\semcolor \calA$ is combinatory complete and extensional with respect to $\semcolor \IA$.
\end{proposition}\label{thm:operad}
While Proposition \ref{thm:operad} can be shown by direct calculation, it is much easier to
make use of the notion of arities.  
Following our observation on arities on the closed $\lambda$-terms (Proposition \ref{prop:arity-m-n}), we define:
\begin{definition}
An element $a$ of an extensional $\BIdot$-algebra is
said to be of arity $m\rightarrow n$
when
$a^\bullet\circ\BB^{m+1}=(\BB\,a)\circ\BB^n$
holds.
\end{definition}
It follows that $a:m\rightarrow 1$ iff $a\in\IA(m)$.
We shall note that the last three axioms of extensional $\BIdot$-algebras say 
$\semcolor \BB:2\rightarrow 1$ (B$\bullet$), 
$\semcolor \II:0\rightarrow0$ (I$\bullet$) and
$\semcolor a^\bullet:0\rightarrow 1$ ($\bullet\bullet$) respectively.

\begin{lemma} \label{lem:compositionality}
The following hold in extensional $\BIdot$-algebras.
\begin{enumerate}
\item 
$\semcolor \BB:2\rightarrow 1$, 
$\semcolor \II:0\rightarrow0$ and
$\semcolor a^\bullet:0\rightarrow 1$.
\item
If $a:l\rightarrow m$
and $b:m\rightarrow n$, then
$a\circ b:l\rightarrow n$.
\item 
If $a:m\rightarrow n$, then
$\BB\,a:m+1\rightarrow n+1$. 
Moreover, $\BB\,\II=\II$ and $\BB\,(a\circ b)=(\BB\,a)\circ(\BB\,b)$ hold.
\item 
If $a:m\rightarrow n$, then
$a:m+1\rightarrow n+1$.
\item For $a:m\rightarrow n$ and $b$, $(\BB^m\,b)\circ a = a\circ (\BB^n\,b)$ holds.
\end{enumerate}
\end{lemma}
From this lemma, Proposition \ref{thm:operad} easily follows.
Moreover, using this notion of arity, we can define a PRO (strict monoidal
category whose objects are generated from a single object) of 
extensional $\BIdot$-algebras, into which the internal operad fully faithfully embeds.

\begin{theorem} \label{thm:pro}
For any extensional $\BIdot$-algebra $\calA$,
we have a PRO $\calC_{\calA}$ whose arrows from $m$ to $n$ are $\calA$'s 
elements of arity $m\rightarrow n$. In particular, we have 
$\calC_\calA(m,1)=\IA(m)$.
\end{theorem}
As an immediate corollary, we have a sort of Scott's theorem:
\begin{corollary} \label{cor:scott}
For any extensional $\BIdot$-algebra $\calA$,
there exists a monoidal closed category $\mathcal{D}$ with an object $U$ such that
$U$ is isomorphic to the internal hom $[U,U]$ and the induced 
extensional $\BIdot$-algebra $\mathcal{D}(I,U)$ is isomorphic to $\calA$.
\end{corollary}
Indeed, we may take the presheaf category $\mathbf{Set}^{\calC_\calA^\mathrm{op}}$ (monoidal cocompletion of $\calC_\calA$)
as $\mathcal{D}$
and let $U=\calC_\calA(-,1)$.

In Section \ref{sec:variations}, we will consider symmetric, braided and cartesian cases. For these cases,
Theorem \ref{thm:pro} can be refined as follows: $\calC_{\calA}$ is a PROP
(strict symmetric monoidal
category whose objects are generated from a single object) for the symmetric case,
a PROB (strict braided monoidal
category whose objects are generated from a single object) for the braided case,
and a Lawvere theory for the cartesian case. The appropriate variation of Corollary \ref{cor:scott} also holds for
each case, where $\mathcal{D}$ is symmetric, braided or cartesian, respectively. 

\subsection{Internal Operads vs Planar Polynomials}

There exists a homomorphism $\semcolor F$ of closed operads from 
the internal operad $\semcolor \IA$ to the operad $\semcolor \PA$ of planar polynomials
sending 
$\semcolor f\in\IA(n)$ to
$\semcolor Ff\in\PA(n)$ (hence $\semcolor Ff:\calA^n\rightarrow\calA$) by
$$
{\semcolor 
(Ff)(a_1,\dots,a_n)
=
f\,\II\,a_1\,\dots\,a_n.}
$$

$\semcolor F$ does not have to be faithful. 
As a counterexample, let $\semcolor \mathcal{A}$ be the extensional
$\BIdot$-algebra of closed  terms
of the {\em braided}  $\lambda$-calculus \cite{Has22}
modulo
$\beta\eta$-equality, with $\semcolor \BB\equiv\lambda fxy.f\,(x\,y)$,
$\semcolor \II\equiv\lambda x.x$ and $\semcolor M^\bullet\equiv\lambda f.f\,M$. 
The following two braided terms (in the syntax of \cite{Has22})
$$
{\semcolor 
M^+\equiv\lambda fxy.
\left[
\begin{picture}(30,20)(-5,-2)
\thicklines
\qbezier(0,0)(5,0)(10,5)
\qbezier(10,5)(15,10)(20,10)
\qbezier(0,10)(5,10)(7,8)
\qbezier(13,2)(15,0)(20,0)
\qbezier(0,-10)(0,-10)(20,-10)
\put(-5,10){\makebox(0,0){\footnotesize$y$}}
\put(-5,0){\makebox(0,0){\footnotesize$x$}}
\put(-5,-10){\makebox(0,0){\footnotesize$f$}}
\put(25,10){\makebox(0,0){\footnotesize$x$}}
\put(25,0){\makebox(0,0){\footnotesize$y$}}
\put(25,-10){\makebox(0,0){\footnotesize$f$}}
\end{picture}
\right]
(f\,(y\,x))
~~~~~
M^-\equiv\lambda fxy.
\left[
\begin{picture}(30,20)(-5,-2)
\thicklines
\qbezier(0,10)(5,10)(10,5)
\qbezier(10,5)(15,0)(20,0)
\qbezier(0,0)(5,0)(7,2)
\qbezier(13,8)(15,10)(20,10)
\qbezier(0,-10)(0,-10)(20,-10)
\put(-5,10){\makebox(0,0){\footnotesize$y$}}
\put(-5,0){\makebox(0,0){\footnotesize$x$}}
\put(-5,-10){\makebox(0,0){\footnotesize$f$}}
\put(25,10){\makebox(0,0){\footnotesize$x$}}
\put(25,0){\makebox(0,0){\footnotesize$y$}}
\put(25,-10){\makebox(0,0){\footnotesize$f$}}
\end{picture}
\right]
(f\,(y\,x))
}
$$
give two distinct elements of $\semcolor \IA(2)$. 
However, $\semcolor FM^+$ and $\semcolor FM^-$ are the same map sending $\semcolor (a_1,a_2)$ to $\semcolor a_2\,a_1$,
thus the information on braids is lost in $\semcolor \CA(2)$.
(In fact, while $\semcolor \PA$ is a closed planar operad, 
it is {\em not} a braided operad. On the other hand, in Section \ref{sec:variations}
we will see that
$\semcolor \IA$ is a closed braided operad.)

\subsection{The Canonicity of Internal Operads}

In fact, the internal operad is the canonical -- initial -- one
among the closed operads corresponding to an extensional $\BIdot$-algebra.

\begin{proposition}
Let $\calA$ be an extensional $\BIdot$-algebra and 
$\calP$ a closed operad such that $\calP(0)\cong\calA$.
Then there is a unique homomorphism of closed operads from
$\IA$ to $\calP$. 
\end{proposition}
Explicitly, the homomorphism from
$\semcolor\IA$ to $\semcolor\calP$ 
sends (assuming $\semcolor\calP(0)=\calA$ for simplicity)
$t\semcolor \in\IA(m)$ to
$\semcolor\mathbf{app}_\calP(\dots(\mathbf{app}_\calP(t\,\II,\id_\calP),\id_\calP)\dots,\id_\calP)\in \calP(m).$
More succinctly, we have
\begin{theorem}
The internal operad construction $\calA\mapsto \IA$ gives a
left adjoint to the functor from the category of closed operads (and operad homomorphisms preserving the closed structure) to
that of extensional $\BIdot$-algebras (and maps preserving the $\BIdot$-algebra structure)
sending a closed operad 
$\calP$ to an extensional $\BIdot$-algebra $\calP(0)$.
\end{theorem}

\section{Variations}
\label{sec:variations}

\subsection{Extensional $\BCI$-algebras and Closed Symmetric Operads}

\newsavebox{\smallBC}
\savebox{\smallBC}{%
\unitlength=.7pt
\begin{picture}(30,40)(0,0)\thicklines
\put(15,40){\oval(30,60)}
\put(15,10){\vector(0,-1){15}}
\qbezier(0,55)(12,55)(15,47.5)\qbezier(15,47.5)(18,40)(30,40)
\qbezier(0,40)(12,40)(15,47.5)\qbezier(15,47.5)(18,55)(30,55)
\qbezier(0,25)(0,25)(30,25)
\put(0,25){\lamnode}
\put(0,40){\lamnode}
\put(0,55){\lamnode}
\put(30,25){\appnode}
\put(30,40){\appnode}
\put(30,55){\appnode}
\put(15,10){\lamnode}
\end{picture}
}%
\comment{
\savebox{\smallBC}{%
\unitlength=.8pt
\begin{picture}(30,40)(0,0)\thicklines
\put(15,25){\circle{30}}
\put(15,10){\vector(0,-1){15}}
\qbezier(4,35)(12,35)(15,30)\qbezier(15,30)(18,25)(30,25)
\qbezier(0,25)(12,25)(15,30)\qbezier(15,30)(18,35)(26,35)
\qbezier(4,15)(4,15)(26,15)
\put(4,15){\tinylamnode}
\put(26,15){\tinyappnode}
\put(0,25){\tinylamnode}
\put(30,25){\tinyappnode}
\put(4,35){\tinylamnode}
\put(26,35){\tinyappnode}
\put(15,9){\tinylamnode}
\end{picture}
}%
}
\newsavebox{\smallC}
\savebox{\smallC}{%
\unitlength=.7pt
\begin{picture}(30,40)(0,0)\thicklines
\put(15,40){\oval(30,60)}
\put(15,10){\vector(0,-1){15}}
\qbezier(0,25)(12,25)(15,32.5)\qbezier(15,32.5)(18,40)(30,40)
\qbezier(0,40)(12,40)(15,32.5)\qbezier(15,32.5)(18,25)(30,25)
\qbezier(0,55)(0,55)(30,55)
\put(0,25){\lamnode}
\put(0,40){\lamnode}
\put(0,55){\lamnode}
\put(30,25){\appnode}
\put(30,40){\appnode}
\put(30,55){\appnode}
\put(15,10){\lamnode}
\end{picture}
}%
\comment{
\savebox{\smallC}{%
\unitlength=.8pt
\begin{picture}(30,40)(0,0)\thicklines
\put(15,25){\circle{30}}
\qbezier(15,10)(15,10)(15,0)
\qbezier(4,15)(12,15)(15,20)\qbezier(15,20)(18,25)(30,25)
\qbezier(0,25)(12,25)(15,20)\qbezier(15,20)(18,15)(26,15)
\qbezier(4,35)(4,35)(26,35)
\put(4,15){\tinylamnode}
\put(26,15){\tinyappnode}
\put(0,25){\tinylamnode}
\put(30,25){\tinyappnode}
\put(4,35){\tinylamnode}
\put(26,35){\tinyappnode}
\put(15,9){\tinylamnode}
\end{picture}
}%
}

An {\em extensional $\BCI$-algebra} is an applicative structure with
elements $\semcolor \BB$, $\semcolor \CC$ and $\semcolor \II$ satisfying the following axioms.
{\semcolor 
$$
\begin{array}{|rcll|}
\hline
\BB\,a\,b\,c &=& a\,(b\,c) & (B)\\
\CC\,a\,b\,c &=& a\,c\,b & (C)\\
\II\,a &=& a & (I)\\ 
\hline
\BB\,\II &=& \II  & (\lambda)\\
\CC\,\BB\,\II &=& \II & (\rho)\\
(\BB\,\BB)\circ\BB &=& (\CC\,\BB\,\BB)\circ(\BB\circ\BB) & (\alpha)\\
\hline
\CC\circ\CC &=& \II & (\cox_1)\\
(\BB\,\CC)\circ(\BB\circ\BB) &=& (\CC\,\BB\,\CC)\circ(\BB\circ\BB) & (\cox_2)\\
(\BB\,\CC)\circ(\CC\circ(\BB\,\CC)) &=& 
\CC\circ((\BB\,\CC)\circ\CC) & (\cox_3)\\
(\BB\,\BB)\circ\CC &=& \CC\circ((\BB\,\CC)\circ\BB) &(bc)\\
\hline
\end{array}
$$
}\noindent
These axioms first appeared in our previous work \cite{Has22}.
They are {\em  chosen} so that the internal operad construction gives rise to a
{\keywordcolor closed symmetric operad}: $(\lambda)$, $(\rho)$ and $(\alpha)$ are for the unit law and associativity of the composition, while 
$(\cox_{1,2,3})$ are the axioms of symmetric groups 
and  $(bc)$ is the equivariance of symmetry with respect to the  application ---
also it is the Reidemeister move IV for trivalent graphs.

Recall that the {\em symmetric group} on $n$ elements is generated by the 
adjacent transpositions $\semcolor \sigma _{i}=(i,i+1)$ ($\semcolor 1\leq i\leq n-1$) subject to the 
following relations (known as {\keywordcolor Coxeter relations}):
$$
\sigma_i^2 = e ,~~~~
\sigma_i\sigma_j = \sigma_j\sigma_i ~ (j<i-1),~~~~ 
\sigma_{i+1}\sigma_i\sigma_{i+1} = \sigma_i\sigma_{i+1}\sigma_i.
$$
\comment{
$${\semcolor 
\begin{array}{rcll}
\sigma_i^2 &=& e \\
\sigma_i\sigma_j &=& \sigma_j\sigma_i & (|i-j|\geq 2) \\
\sigma_{i+1}\sigma_i\sigma_{i+1} &=& \sigma_i\sigma_{i+1}\sigma_i
\end{array}}
$$
}
The axioms ($\cox_1$), ($\cox_2$) and ($\cox_3$) correspond
to these axioms of symmetric groups. $(\cox_1)$ can be depicted as
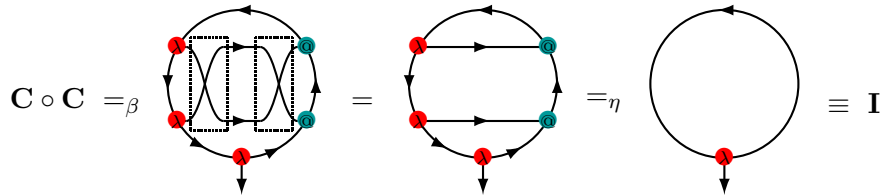
\begin{center}
\unitlength=.7pt
\begin{picture}(150,110)(-50,0)
\put(-40,50){\makebox(0,0){$\CC\circ\CC~=_\beta$}}
\thicklines
\put(50,60){\circle{80}}
\put(50,20){\vector(0,-1){20}}
\put(47,100){\vector(-1,0){1}}
\put(10,57){\vector(0,-1){1}}
\put(30,25.5){\vector(2,-1){1}}
\put(70.5,25.5){\vector(2,1){1}}
\put(90,63){\vector(0,1){1}}
\put(53,80){\vector(1,0){1}}
\put(53,40){\vector(1,0){1}}
\qbezier(15,80)(15,80)(20,80)\qbezier(20,80)(25,80)(30,60)\qbezier(30,60)(35,40)(40,40)\qbezier(40,40)(50,40)(60,40)\qbezier(60,40)(65,40)(70,60)\qbezier(70,60)(75,80)(80,80)\qbezier(80,80)(85,80)(85,80)
\qbezier(15,40)(15,40)(20,40)\qbezier(20,40)(25,40)(30,60)\qbezier(30,60)(35,80)(40,80)\qbezier(40,80)(50,80)(60,80)\qbezier(60,80)(65,80)(70,60)\qbezier(70,60)(75,40)(80,40)\qbezier(80,40)(85,40)(85,40)
\put(50,20){\lamnode}
\put(15,80){\lamnode}
\put(15,40){\lamnode}
\put(85,80){\appnode}
\put(85,40){\appnode}
\put(22.5,35){\dashbox(20,50){}}
\put(57.5,35){\dashbox(20,50){}}
\end{picture}
\begin{picture}(20,110)
\put(10,50){\makebox(0,0){$=$}}
\end{picture}
\begin{picture}(100,110)
\comment{%
\put(0,80){\line(1,0){100}}
\put(0,60){\line(1,0){100}}
\put(0,40){\line(1,0){100}}
\put(20,0){\line(0,1){100}}
\put(40,0){\line(0,1){100}}
\put(60,0){\line(0,1){100}}
\put(80,0){\line(0,1){100}}
}
\thicklines
\put(50,60){\circle{80}}
\put(50,20){\vector(0,-1){20}}
\put(47,100){\vector(-1,0){1}}
\put(10,57){\vector(0,-1){1}}
\put(30,25.5){\vector(2,-1){1}}
\put(70.5,25.5){\vector(2,1){1}}
\put(90,63){\vector(0,1){1}}
\put(53,80){\vector(1,0){1}}
\put(53,40){\vector(1,0){1}}
\qbezier(15,80)(15,80)(85,80)
\qbezier(15,40)(15,40)(85,40)
\put(50,20){\lamnode}
\put(15,80){\lamnode}
\put(15,40){\lamnode}
\put(85,80){\appnode}
\put(85,40){\appnode}
\end{picture}
\begin{picture}(20,110)
\put(10,50){\makebox(0,0){$=_\eta$}}
\end{picture}
\begin{picture}(120,110)
\put(120,50){\makebox(0,0){$\equiv~\II$}}
\comment{%
\put(0,80){\line(1,0){100}}
\put(0,60){\line(1,0){100}}
\put(0,40){\line(1,0){100}}
\put(20,0){\line(0,1){100}}
\put(40,0){\line(0,1){100}}
\put(60,0){\line(0,1){100}}
\put(80,0){\line(0,1){100}}
}
\thicklines
\put(50,60){\circle{80}}
\put(50,20){\vector(0,-1){20}}
\put(47,100){\vector(-1,0){1}}
\put(50,20){\lamnode}
\end{picture}
\end{center}
which amounts to the axiom $\semcolor \sigma_i\sigma_i=e$ 
of the symmetric  groups.

$(\cox_2)$ is equivalent to say that $\semcolor \CC$ is of arity $\semcolor 2\rightarrow 2$,
and expresses the following exchange law,
which corresponds to the axiom 
$\semcolor \sigma_i\sigma_j=\sigma_j\sigma_i$
($\semcolor j<i-1$) of symmetric groups:
\begin{center}
\unitlength=.7pt
\begin{picture}(100,140)(0,0)
\put(-65,70){\makebox(0,0){$(\BB\,(\BB\,L))\circ\CC~=_\beta$} }
\thicklines
\put(50,90){\circle{100}}
\rgbfbox{25}{95}{20}{30}{1}{1}{0}{}%

\qbezier(10,120)(10,120)(25,120) \qbezier(45,120)(65,120)(90,120)
\qbezier(0,100)(0,100)(25,100) \qbezier(45,100)(100,100)(100,100)
\qbezier(0,80)(0,80)(55,80) \qbezier(75,80)(100,80)(100,80)
\qbezier(10,60)(10,60)(55,60) \qbezier(75,60)(90,60)(90,60)
\qbezier(55,80)(60,80)(65,70) \qbezier(65,70)(70,60)(75,60)
\qbezier(55,60)(60,60)(65,70) \qbezier(65,70)(70,80)(75,80)
\put(15,120){\vector(1,0){10}}
\put(15,100){\vector(1,0){10}}
\put(15,80){\vector(1,0){10}}

\put(15,60){\vector(1,0){10}}

\put(85,120){\vector(1,0){1}}
\put(85,100){\vector(1,0){1}}
\put(85,80){\vector(1,0){1}}
\put(85,60){\vector(1,0){1}}
\put(50,40){\vector(0,-1){30}}
\put(50,140){\vector(-1,0){1}}
\put(4,110){\vector(-1,-2){1}}
\put(0,87){\vector(0,-1){1}}
\put(5,68){\vector(1,-2){1}}

\put(96,110){\vector(-1,2){1}}
\put(100,93){\vector(0,1){1}}
\put(96,70){\vector(1,2){1}}
\put(30,44){\vector(2,-1){1}}
\put(77,48){\vector(2,1){1}}
\put(10,120){\lamnode}
\put(0,100){\lamnode}
\put(0,80){\lamnode}
\put(10,60){\lamnode}
\put(50,40){\lamnode}
\put(90,120){\appnode}
\put(100,100){\appnode}
\put(100,80){\appnode}
\put(90,60){\appnode}
\put(22.5,52.5){\dashbox(25,75){}}
\put(52.5,52.5){\dashbox(25,75){}}
\end{picture}
\begin{picture}(40,140)(0,0)
\put(20,70){\makebox(0,0){$=$}}
\end{picture}
\begin{picture}(100,140)(0,0)
\put(165,70){\makebox(0,0){$=_\beta~\CC\circ(\BB\,(\BB\,L))$}}
\thicklines
\put(50,90){\circle{100}}
\rgbfbox{55}{95}{20}{30}{1}{1}{0}{}%

\qbezier(10,120)(10,120)(55,120) \qbezier(75,120)(90,120)(90,120)
\qbezier(0,100)(0,100)(55,100) \qbezier(75,100)(100,100)(100,100)
\qbezier(0,80)(0,80)(25,80) \qbezier(45,80)(100,80)(100,80)
\qbezier(10,60)(10,60)(25,60) \qbezier(45,60)(90,60)(90,60)
\qbezier(25,80)(30,80)(35,70) \qbezier(35,70)(40,60)(45,60)
\qbezier(25,60)(30,60)(35,70) \qbezier(35,70)(40,80)(45,80)
\put(15,120){\vector(1,0){10}}
\put(15,100){\vector(1,0){10}}
\put(15,80){\vector(1,0){10}}

\put(15,60){\vector(1,0){10}}

\put(85,120){\vector(1,0){1}}
\put(85,100){\vector(1,0){1}}
\put(85,80){\vector(1,0){1}}
\put(85,60){\vector(1,0){1}}
\put(50,40){\vector(0,-1){30}}
\put(50,140){\vector(-1,0){1}}
\put(4,110){\vector(-1,-2){1}}
\put(0,87){\vector(0,-1){1}}
\put(5,68){\vector(1,-2){1}}

\put(96,110){\vector(-1,2){1}}
\put(100,93){\vector(0,1){1}}
\put(96,70){\vector(1,2){1}}
\put(30,44){\vector(2,-1){1}}
\put(77,48){\vector(2,1){1}}
\put(10,120){\lamnode}
\put(0,100){\lamnode}
\put(0,80){\lamnode}
\put(10,60){\lamnode}
\put(50,40){\lamnode}
\put(90,120){\appnode}
\put(100,100){\appnode}
\put(100,80){\appnode}
\put(90,60){\appnode}
\put(22.5,52.5){\dashbox(25,75){}}
\put(52.5,52.5){\dashbox(25,75){}}
\end{picture}
\end{center}

Finally, $(\cox_3)$ is
\begin{center}
\unitlength=.7pt
\begin{picture}(140,110)(-40,0)
\put(-70,50){\makebox(0,0){$(\BB\,\CC)\circ\CC\circ(\BB\,\CC)~=_\beta$} }
\thicklines
\put(50,60){\circle{80}}
\put(50,20){\vector(0,-1){20}}
\put(50,100){\vector(-1,0){1}}
\put(11,69){\vector(-1,-4){1}}
\put(11,51){\vector(1,-4){1}}
\put(30,25.5){\vector(2,-1){1}}
\put(70.5,25.5){\vector(2,1){1}}
\put(89,51){\vector(1,4){1}}
\put(89,69){\vector(-1,4){1}}
\put(30,40){\vector(1,0){1}}
\put(50,80){\vector(1,0){1}}
\put(70,40){\vector(1,0){1}}
\qbezier(15,80)(15,80)(20,80)\qbezier(20,80)(25,80)(30,70)\qbezier(30,70)(35,60)(40,60)\qbezier(40,60)(45,60)(50,50)\qbezier(50,50)(55,40)(60,40)\qbezier(60,40)(70,40)(85,40)
\qbezier(10,60)(15,60)(20,60)\qbezier(20,60)(25,60)(30,70)\qbezier(30,70)(35,80)(40,80)\qbezier(40,80)(50,80)(60,80)\qbezier(60,80)(65,80)(70,70)\qbezier(70,70)(75,60)(80,60)\qbezier(80,60)(85,60)(90,60)
\qbezier(15,40)(15,40)(40,40)\qbezier(40,40)(45,40)(50,50)\qbezier(50,50)(55,60)(60,60)\qbezier(60,60)(65,60)(70,70)\qbezier(70,70)(75,80)(80,80)\qbezier(80,80)(85,80)(85,80)
\put(50,20){\lamnode}
\put(15,80){\lamnode}
\put(10,60){\lamnode}
\put(15,40){\lamnode}
\put(85,80){\appnode}
\put(90,60){\appnode}
\put(85,40){\appnode}
\put(22.5,35){\dashbox(15,50){}}
\put(42.5,35){\dashbox(15,50){}}
\put(62.5,35){\dashbox(15,50){}}
\end{picture}
\begin{picture}(20,110)
\put(10,50){\makebox(0,0){$=$}}
\end{picture}
\begin{picture}(100,110)
\put(160,50){\makebox(0,0){$=_\beta~\CC\circ(\BB\,\CC)\circ\CC$}}
\comment{%
\put(0,80){\line(1,0){100}}
\put(0,60){\line(1,0){100}}
\put(0,40){\line(1,0){100}}
\put(20,0){\line(0,1){100}}
\put(40,0){\line(0,1){100}}
\put(60,0){\line(0,1){100}}
\put(80,0){\line(0,1){100}}
}%
\thicklines
\put(50,60){\circle{80}}
\put(50,20){\vector(0,-1){20}}
\put(50,100){\vector(-1,0){1}}
\put(11,69){\vector(-1,-4){1}}
\put(11,51){\vector(1,-4){1}}
\put(30,25.5){\vector(2,-1){1}}
\put(70.5,25.5){\vector(2,1){1}}
\put(89,51){\vector(1,4){1}}
\put(89,69){\vector(-1,4){1}}
\put(30,80){\vector(1,0){1}}
\put(50,40){\vector(1,0){1}}
\put(70,80){\vector(1,0){1}}
\qbezier(15,80)(15,80)(40,80)\qbezier(40,80)(45,80)(50,70)\qbezier(50,70)(55,60)(60,60)\qbezier(60,60)(65,60)(70,50)\qbezier(70,50)(75,40)(80,40)\qbezier(80,40)(85,40)(85,40)
\qbezier(10,60)(15,60)(20,60)\qbezier(20,60)(25,60)(30,50)\qbezier(30,50)(35,40)(40,40)\qbezier(40,40)(50,40)(60,40)\qbezier(60,40)(65,40)(70,50)\qbezier(70,50)(75,60)(80,60)\qbezier(80,60)(85,60)(90,60)
\qbezier(15,40)(15,40)(20,40)\qbezier(20,40)(25,40)(30,50)\qbezier(30,50)(35,60)(40,60)\qbezier(40,60)(45,60)(50,70)\qbezier(50,70)(55,80)(60,80)\qbezier(60,80)(70,80)(85,80)
\put(50,20){\lamnode}
\put(15,80){\lamnode}
\put(10,60){\lamnode}
\put(15,40){\lamnode}
\put(85,80){\appnode}
\put(90,60){\appnode}
\put(85,40){\appnode}
\put(22.5,35){\dashbox(15,50){}}
\put(42.5,35){\dashbox(15,50){}}
\put(62.5,35){\dashbox(15,50){}}
\end{picture}
\end{center}
which is the axiom $\semcolor \sigma_{i+1}\sigma_i\sigma_{i+1}=\sigma_i\sigma_{i+1}\sigma_i$
of the symmetric  groups.

On the other hand, the axiom $(bc)$ is
\begin{center}
\unitlength=.75pt
\begin{picture}(100,110)
\put(-45,50){\makebox(0,0){$(\BB\,\BB)\circ\CC~=_\beta$} }
\thicklines
\put(50,60){\circle{80}}
\put(50,20){\vector(0,-1){20}}
\put(50,100){\vector(-1,0){1}}
\put(11,69){\vector(-1,-4){1}}
\put(11,51){\vector(1,-4){1}}
\put(30,25.5){\vector(2,-1){1}}
\put(70.5,25.5){\vector(2,1){1}}
\put(89,51){\vector(1,4){1}}
\put(89,69){\vector(-1,4){1}}
\put(30,40){\vector(1,0){1}}
\put(30,60){\vector(1,0){1}}
\put(30,80){\vector(1,0){1}}
\put(80,50){\vector(1,0){1}}
\qbezier(15,80)(15,80)(30,80)\qbezier(30,80)(35,80)(40,70)\qbezier(40,70)(40,70)(50,70)\qbezier(50,70)(55,70)(60,60)\qbezier(60,60)(65,50)(70,50)\qbezier(70,50)(70,50)(88,50)
\qbezier(10,60)(15,60)(30,60)\qbezier(30,60)(35,60)(40,70)
\qbezier(15,40)(15,40)(50,40)\qbezier(50,40)(55,40)(60,50)\qbezier(60,50)(60,50)(70,70)\qbezier(70,70)(75,80)(80,80)\qbezier(80,80)(85,80)(85,80)
\put(50,20){\lamnode}
\put(15,80){\lamnode}
\put(10,60){\lamnode}
\put(15,40){\lamnode}
\put(85,80){\appnode}
\put(40,70){\appnode}
\put(88,50){\appnode}
\put(22.5,35){\dashbox(25,50){}}
\put(52.5,35){\dashbox(25,50){}}
\end{picture}
\begin{picture}(20,110)
\put(10,50){\makebox(0,0){$=$}}
\end{picture}
\begin{picture}(100,110)
\put(160,50){\makebox(0,0){$=_\beta~\CC\circ(\BB\,\CC)\circ\BB$} }
\thicklines
\put(50,60){\circle{80}}
\put(50,20){\vector(0,-1){20}}
\put(50,100){\vector(-1,0){1}}
\put(11,69){\vector(-1,-4){1}}
\put(11,51){\vector(1,-4){1}}
\put(30,25.5){\vector(2,-1){1}}
\put(70.5,25.5){\vector(2,1){1}}
\put(83,50){\vector(1,0){1}}
\put(89,69){\vector(-1,4){1}}
\put(30,80){\vector(1,0){1}}
\put(50,40){\vector(1,0){1}}
\put(70,80){\vector(1,0){1}}
\qbezier(15,80)(15,80)(40,80)\qbezier(40,80)(45,80)(50,70)\qbezier(50,70)(55,60)(60,60)\qbezier(60,60)(65,60)(70,50)
\qbezier(10,60)(15,60)(20,60)\qbezier(20,60)(25,60)(30,50)\qbezier(30,50)(35,40)(40,40)\qbezier(40,40)(50,40)(60,40)\qbezier(60,40)(65,40)(70,50)
\qbezier(15,40)(15,40)(20,40)\qbezier(20,40)(25,40)(30,50)\qbezier(30,50)(35,60)(40,60)\qbezier(40,60)(45,60)(50,70)\qbezier(50,70)(55,80)(60,80)\qbezier(60,80)(70,80)(85,80)
\qbezier(70,50)(70,50)(88,50)
\put(50,20){\lamnode}
\put(15,80){\lamnode}
\put(10,60){\lamnode}
\put(15,40){\lamnode}
\put(85,80){\appnode}
\put(70,50){\appnode}
\put(88,50){\appnode}
\put(22.5,35){\dashbox(15,50){}}
\put(42.5,35){\dashbox(15,50){}}
\put(62.5,35){\dashbox(15,50){}}
\end{picture}
\end{center}
which amounts to the {\kc Reidemeister move IV} \cite{Kau89,Yet89} 
 for knotted graphs:
\begin{center}
\unitlength=.6pt
\begin{picture}(70,50)(15,35)
\thicklines
\qbezier(15,80)(15,80)(30,80)\qbezier(30,80)(35,80)(40,70)\qbezier(40,70)(40,70)(50,70)\qbezier(50,70)(55,70)(60,60)\qbezier(60,60)(65,50)(70,50)\qbezier(70,50)(70,50)(85,50)
\qbezier(15,60)(15,60)(30,60)\qbezier(30,60)(35,60)(40,70)
\qbezier(15,40)(15,40)(50,40)\qbezier(50,40)(55,40)(60,50)
\qbezier(65,60)(70,70)(70,70)\qbezier(70,70)(75,80)(80,80)\qbezier(80,80)(85,80)(85,80)
\end{picture}
\begin{picture}(100,50)(-10,0)
\put(40,29){\makebox(0,0){\footnotesize\scriptsize R-IV}}
\put(40,20){\vector(1,0){40}}
\put(40,20){\vector(-1,0){40}}
\end{picture}
\begin{picture}(70,50)(15,35)
\thicklines
\qbezier(15,80)(15,80)(40,80)\qbezier(40,80)(45,80)(50,70)\qbezier(50,70)(55,60)(60,60)\qbezier(60,60)(65,60)(70,50)
\qbezier(15,60)(15,60)(20,60)\qbezier(20,60)(25,60)(30,50)\qbezier(30,50)(35,40)(40,40)\qbezier(40,40)(50,40)(60,40)\qbezier(60,40)(65,40)(70,50)
\qbezier(15,40)(15,40)(20,40)\qbezier(20,40)(25,40)(28,46)\qbezier(32,54)(35,60)(40,60)\qbezier(40,60)(45,60)(48,66)\qbezier(52,74)(55,80)(60,80)\qbezier(60,80)(70,80)(85,80)
\qbezier(70,50)(70,50)(85,50)
\end{picture}
~~~~~~
\begin{picture}(70,50)(15,35)
\thicklines
\qbezier(15,80)(15,80)(30,80)\qbezier(30,80)(35,80)(40,70)\qbezier(40,70)(40,70)(50,70)\qbezier(50,70)(55,70)(60,60)\qbezier(64,52)(65,50)(70,50)\qbezier(70,50)(70,50)(85,50)
\qbezier(15,60)(15,60)(30,60)\qbezier(30,60)(35,60)(40,70)
\qbezier(15,40)(15,40)(50,40)\qbezier(50,40)(55,40)(60,50)
\qbezier(60,50)(70,70)(70,70)\qbezier(70,70)(75,80)(80,80)\qbezier(80,80)(85,80)(85,80)
\end{picture}
\begin{picture}(100,50)(-10,0)
\put(40,29){\makebox(0,0){\footnotesize\scriptsize R-IV}}
\put(40,20){\vector(1,0){40}}
\put(40,20){\vector(-1,0){40}}
\end{picture}
\begin{picture}(70,50)(15,35)
\thicklines
\qbezier(15,80)(15,80)(40,80)\qbezier(40,80)(45,80)(48,74)\qbezier(52,66)(55,60)(60,60)\qbezier(60,60)(65,60)(70,50)
\qbezier(15,60)(15,60)(20,60)\qbezier(20,60)(25,60)(28,54)\qbezier(32,46)(35,40)(40,40)\qbezier(40,40)(50,40)(60,40)\qbezier(60,40)(65,40)(70,50)
\qbezier(15,40)(15,40)(20,40)\qbezier(20,40)(25,40)(30,50)\qbezier(30,50)(35,60)(40,60)\qbezier(40,60)(45,60)(50,70)\qbezier(50,70)(55,80)(60,80)\qbezier(60,80)(70,80)(85,80)
\qbezier(70,50)(70,50)(85,50)
\end{picture}
\end{center}

A {\em symmetric operad} is an operad equipped with actions of symmetric groups
satisfying equivariance conditions (see the case of braid groups below).

\begin{lemma}
An extensional $\BCI$-algebra is also an 
extensional $\BIdot$-algebra with $\semcolor a^\bullet=\CC\,\II\,a$.
\end{lemma}

\begin{proposition}
For an extensional $\BCI$-algebra $\semcolor \calA$, $\semcolor \IA$ is a
closed symmetric operad s.t. $\semcolor \IA(0)\cong\calA$.
\end{proposition}

\begin{theorem} \cite{Has22}
Extensional $\semcolor \BCI$-algebras are sound and complete for
the linear $\lambda\beta\eta$-calculus.
\end{theorem}

\begin{theorem}
The internal operad construction $\calA\mapsto \IA$ is
left adjoint to the functor from the category of closed symmetric operads to
that of extensional $\BCI$-algebras sending $\calP$ to $\calP(0)$.
\end{theorem}

\subsection{Extensional $\BCpmI$-algebras and Closed Braided Operads}

{\em Extensional $\BCpmI$-algebras} are a refinement of extensional 
$\BCI$-algebras
in which the $\CC$-combinator is replaced by
the combinators $\CC^+$, $\CC^-$ for positive and negative {\kc braids}:
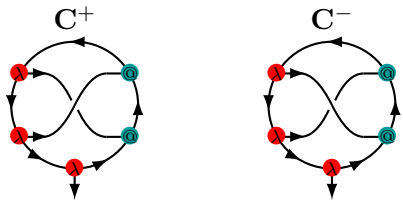
\begin{center}
\unitlength=.6pt
\begin{picture}(100,120)
\put(50,115){\makebox(0,0){$\CC^+$}}
\thicklines
\put(50,60){\circle{80}}
\put(50,20){\vector(0,-1){20}}
\put(47,100){\vector(-1,0){1}}
\put(10,57){\vector(0,-1){1}}
\put(30,25.5){\vector(2,-1){1}}
\put(70.5,25.5){\vector(2,1){1}}
\put(90,63){\vector(0,1){1}}
\put(33,80){\vector(1,0){1}}
\put(33,40){\vector(1,0){1}}
\qbezier(15,80)(15,80)(30,80)\qbezier(30,80)(40,80)(48,63)\qbezier(52,57)(60,40)(70,40)\qbezier(70,40)(85,40)(85,40)
\qbezier(15,40)(15,40)(30,40)\qbezier(30,40)(40,40)(50,60)\qbezier(50,60)(60,80)(70,80)\qbezier(70,80)(85,80)(85,80)
\put(50,20){\lamnode}
\put(15,80){\lamnode}
\put(15,40){\lamnode}
\put(85,80){\appnode}
\put(85,40){\appnode}
\end{picture}
\begin{picture}(50,100)
\end{picture}
\begin{picture}(100,120)
\put(50,115){\makebox(0,0){$\CC^-$}}
\thicklines
\put(50,60){\circle{80}}
\put(50,20){\vector(0,-1){20}}
\put(47,100){\vector(-1,0){1}}
\put(10,57){\vector(0,-1){1}}
\put(30,25.5){\vector(2,-1){1}}
\put(70.5,25.5){\vector(2,1){1}}
\put(90,63){\vector(0,1){1}}
\put(33,80){\vector(1,0){1}}
\put(33,40){\vector(1,0){1}}
\qbezier(15,80)(15,80)(30,80)\qbezier(30,80)(40,80)(50,60)\qbezier(50,60)(60,40)(70,40)\qbezier(70,40)(85,40)(85,40)
\qbezier(15,40)(15,40)(30,40)\qbezier(30,40)(40,40)(48,57)\qbezier(52,63)(60,80)(70,80)\qbezier(70,80)(85,80)(85,80)
\put(50,20){\lamnode}
\put(15,80){\lamnode}
\put(15,40){\lamnode}
\put(85,80){\appnode}
\put(85,40){\appnode}
\end{picture}
\end{center}
An extensional $\BCpmI$-algebra is an applicative structure 
with elements $\semcolor \BB$, $\semcolor \CC^+$, $\semcolor \CC^-$ and $\semcolor \II$ satisfying the
following axioms.

$${\semcolor 
\begin{array}{|rcll|}
\hline
\BB\,a\,b\,c &=& a\,(b\,c) & (B)\\
\CC^\star\,a\,b\,c &=& a\,c\,b & (C)\\
\II\,a &=& a & (I)\\ 
\hline
\CC^+\,a\,b &=& \CC^-\,a\,b & (C2)\\
\BB\,\II &=& \II  & (\lambda)\\
\CC^\star\,\BB\,\II &=& \II & (\rho)\\
(\BB\,\BB)\circ\BB &=& (\CC^\star\,\BB\,\BB)\circ(\BB\circ\BB) & (\alpha)\\
\hline
\CC^\pm\circ\CC^\mp &=& \II & (\cox_1)\\
(\BB\,\CC^\pm)\circ(\BB\circ\BB) &=& (\CC^\star\,\BB\,\CC^\pm)\circ(\BB\circ\BB) & (\cox_2)\\
(\BB\,\CC^\pm)\circ(\CC^\pm\circ(\BB\,\CC^\pm)) &=& 
\CC^\pm\circ((\BB\,\CC^\pm)\circ\CC^\pm) & (\cox_3)\\
(\BB\,\BB)\circ\CC^\pm &=& \CC^\pm\circ((\BB\,\CC^\pm)\circ\BB) &(bc)\\
\hline
\end{array}}
$$ 
The double signs $\pm$ and $\mp$ in an equation should be taken 
as appropriately linked, while $\star$ indicates an arbitrary choice of $+$ or $-$.
(As we have $(C2)$, assuming just an instance of $\star$ suffices.)

Closed terms of the braided $\lambda$-calculus \cite{Has22} modulo the $\beta\eta$-theory form an extensional $\BCpmI$-algebra. For a non-syntactic example,
for any group $G$, the crossed $G$-set of inifinte binary $G$-labelled trees \cite{Has22} is an  extensional $\BCpmI$-algebra; it is obtained as a reflexive
object in the ribbon category of crossed $G$-sets and suitable relations \cite{Has12}.


\newsavebox{\redbraid}
\savebox{\redbraid}{%
\unitlength=.8pt
\begin{picture}(20,20)(0,0)\thicklines
\qbezier(0,20)(0,12)(10,10)\qbezier(10,10)(20,8)(20,0)
\red\qbezier(20,20)(20,12)(14,11.5)\qbezier(6,8.5)(0,8)(0,0)
\end{picture}}

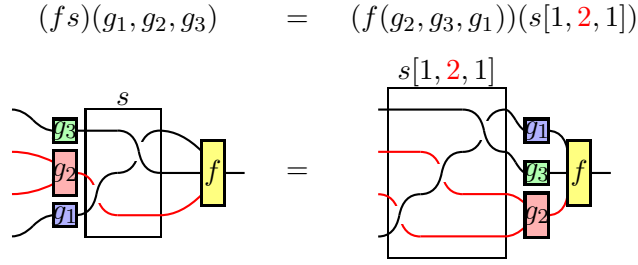
\begin{figure}
\unitlength=.8pt
$$
\begin{array}{ccc}
(fs)(g_1,g_2,g_3)
&=&
(f(g_2,g_3,g_1))(s[1,{\red 2},1])
\\
\\
\begin{picture}(110,80)
\put(35,10){\framebox(35,60){}}
\put(52.5,75){\makebox(0,0){$s$}}
\thicklines
\qbezier(0,10)(5,10)(10,15)\qbezier(10,15)(15,20)(20,20)
{\red\qbezier(0,30)(10,30)(20,35)}%
{\red\qbezier(0,50)(10,50)(20,45)}%
\qbezier(0,70)(5,70)(10,65)\qbezier(10,65)(15,60)(20,60)
\put(30,20){\usebox{\redhbraid}}
\put(50,40){\usebox{\hbraidInv}}
\myline{30}{60}{50}{60}
{\red\myline{50}{20}{70}{20}}%
{\red\qbezier(70,20)(80,20)(90,30)}%
\myline{70}{40}{90}{40}
\qbezier(70,60)(80,60)(90,50)
\myline{100}{40}{110}{40}
\rgbfbox{20}{15}{10}{10}{.7}{.7}{1}{$g_1$}%
\rgbfbox{20}{30}{10}{20}{1}{.7}{.7}{$g_2$}%
\rgbfbox{20}{55}{10}{10}{.7}{1}{.7}{$g_3$}%
\rgbfbox{90}{25}{10}{30}{1}{1}{.5}{$f$}%
\end{picture}
&
\begin{picture}(40,80)
\put(20,40){\makebox(0,0){$=$}}
\end{picture}
&
\begin{picture}(110,80)
\put(5,0){\framebox(55,80){}}
\put(32.5,88){\makebox(0,0){$s[1,{\red 2},1]$}}
\thicklines
\myline{0}{70}{40}{70}
{\red\myline{0}{50}{20}{50}}%
{\red\myline{40}{30}{60}{30}}%
{\red\myline{20}{10}{60}{10}}%
\qbezier(60,70)(62,70)(65,65)\qbezier(65,65)(68,60)(70,60)
\qbezier(60,50)(62,50)(65,45)\qbezier(65,45)(68,40)(70,40)
{\red\qbezier(60,10)(65,10)(70,15)
\qbezier(60,30)(65,30)(70,25)}%
\put(0,10){\usebox{\redhbraid}}
\put(20,30){\usebox{\redhbraid}}
\put(40,50){\usebox{\hbraidInv}}
{\red\qbezier(80,20)(90,20)(90,35)}%
\myline{80}{40}{90}{40}
\qbezier(80,60)(90,60)(90,45)
\myline{100}{40}{110}{40}
\rgbfbox{70}{10}{10}{20}{1}{.7}{.7}{$g_2$}%
\rgbfbox{70}{35}{10}{10}{.7}{1}{.7}{$g_3$}%
\rgbfbox{70}{55}{10}{10}{.7}{.7}{1}{$g_1$}%
\rgbfbox{90}{25}{10}{30}{1}{1}{.5}{$f$}%
\end{picture}
\end{array}
$$
\caption{The equivariance condition}
\label{fig:equivariance}
\end{figure}

A {\em braided operad} \cite{Fie96} 
is an operad equipped with actions of {\em braid groups} \cite{Art25,KT08}
satisfying {\em equivariance conditions} needed for handling substitutions
involving braids. For instance, Figure \ref{fig:equivariance} 
presents an instance of the equivariance condition, which shows that
substituting a term with two free variables ($g_2$) for a variable in a braided term 
($f\,s$) 
involves replacing a strand by
two parallel strands in the braid ($s$).%
\footnote{
Equivariance conditions are also found in the definition of substitutions 
in the braided $\lambda$-calculus \cite{Has22}, though
we were not aware of  the relevance of braided operads 
as of preparing that paper. 
}
For further details see Appendix \ref{sec:braided-operads}.

\begin{lemma}
An extensional $\BCpmI$-algebra is also an 
extensional $\BIdot$-algebra with $\semcolor a^\bullet=\CC^+\,\II\,a$.
\end{lemma}

\begin{proposition}
For an extensional $\BCpmI$-algebra $\semcolor \calA$, $\semcolor \IA$ is a
closed braided operad s.t. $\semcolor \IA(0)\cong\calA$.
\end{proposition}
We shall note that the axiom $(C2)$, which has no counterpart in the axioms of
extensional $\BCI$-algebras, is added for making $\IA$ braided; it amounts to
 an instance of the
equivariance condition: 
$(f\sigma_1)(g,\id)=f(\id,g)=(f\sigma_1^{-1})(g,\id)$ for
$f\in\IA(2)$ and $g\in\IA(0)$, where $\sigma_1$ is the generator of
the braid group $B_2$ of two strands which corresponds to $\CC^+$
and $\sigma_1^{-1}$ is its inverse (corresponding to $\CC^-$).

\begin{theorem} 
Extensional $\BCpmI$-algebras are sound and complete for
the braided $\lambda\beta\eta$-calculus.
\end{theorem}

\begin{theorem}
The internal operad construction $\calA\mapsto \IA$ is
left adjoint to the functor from the category of closed braided operads to
that of extensional $\BCpmI$-algebras sending $\calP$ to $\calP(0)$.
\end{theorem}

\subsection{Extensional $\SK$-algebras and Closed Cartesian Operads}

Instead of $\SK$-algebras, we study {\kc $\BCIWK$-algebras} with 
$\semcolor \WW$ corresponding to $\semcolor \lambda fx.f\,x\,x$ and 
$\semcolor \KK$ corresponding to $\semcolor \lambda fx.f$.
 (It is well known that $\SK$ and $\mathbf{BC(I)WK}$  are equivalent
since Curry's work \cite{Cur30}.) 
\comment{%
A naive graphical language does not work well for the non-linear case 
(for which we need to introduce some sort of boxes for specifying the
scope of $\lambda$-abstractions), but if we stick to our graphical presentation, 
$\WW$ and $\KK$ can be depicted as
\begin{center}\unitlength=.5pt
\begin{picture}(100,120)
\put(50,115){\makebox(0,0){$\WW$}}
\thicklines
\put(50,60){\circle{80}}
\put(50,20){\vector(0,-1){20}}
\put(47,100){\vector(-1,0){1}}
\put(30,25.5){\vector(2,-1){1}}
\put(70.5,25.5){\vector(2,1){1}}
\put(90,63){\vector(0,1){1}}
\put(75,80){\vector(1,0){1}}
\put(75,40){\vector(1,0){1}}
\qbezier(10,60)(15,60)(30,60)\qbezier(30,60)(40,60)(50,50)\qbezier(50,50)(60,40)(70,40)\qbezier(70,40)(85,40)(85,40)
\qbezier(15,60)(15,60)(30,60)\qbezier(30,60)(40,60)(50,70)\qbezier(50,70)(60,80)(70,80)\qbezier(70,80)(85,80)(85,80)
\put(50,20){\lamnode}
\put(10,60){\lamnode}
\put(85,80){\appnode}
\put(85,40){\appnode}
\end{picture}
\begin{picture}(40,100)
\end{picture}
\begin{picture}(100,120)
\put(50,115){\makebox(0,0){$\KK$}}
\thicklines
\put(50,60){\circle{80}}
\put(50,20){\vector(0,-1){20}}
\put(47,100){\vector(-1,0){1}}
\put(10,60){\line(1,0){40}}
\put(40,60){\vector(1,0){1}}
\put(50,60){\makebox(0,0){$\bullet$}}
\put(10,60){\lamnode}
\put(50,20){\lamnode}
\end{picture}
\end{center}
}
An {\kc extensional $\BCIWK$-algebra} is an extensional $\BCI$-algebra
with elements
$\semcolor \WW$ and $\semcolor \KK$ subject to the axioms saying
\begin{itemize}
\item $\semcolor \WW:1\rightarrow 2$ and $\semcolor \KK:1\rightarrow 0$,
\item $\semcolor \WW$ and $\semcolor \KK$ form a co-commutative comonoid, and
\item $\semcolor \BB$ and $\semcolor a^\bullet$ are comonoid morphisms
(the latter implies $\semcolor \WW\,a\,b=a\,b\,b$ and $\semcolor \KK\,a\,b=a$).
\end{itemize}
Explicitly, these axioms can be given as follows.
$$
\begin{array}{|rcll|}
\hline
\WW^\bullet\circ\BB\circ\BB &=& (\BB\,\WW)\circ\BB\circ\BB  & (\WW:1\rightarrow 2)\\
\KK^\bullet\circ\BB\circ\BB &=& \BB\,\KK  & (\KK:1\rightarrow 0)\\
\hline
\WW\circ\KK &=& \II & (\mathit{co\mbox{-}unit})\\
\WW\circ\WW &=& \WW\circ(\BB\,\WW) & (\mathit{co\mbox{-}associativity})\\
\WW\circ\CC &=& \WW & (\mathit{co\mbox{-}commutativity})\\
\hline
\BB\circ\WW &=& (\BB\,\WW)\circ\WW\circ(\BB\,\CC)\circ\BB\circ(\BB\,\BB)~~&
(\BB~\mathit{comonoid~morphism})\\
\BB\circ\KK &=& \KK\circ\KK  & (\BB~\mathit{comonoid~morphism})\\
a^\bullet\circ\WW &=& a^\bullet\circ a^\bullet & (a^\bullet~\mathit{comonoid~morphism})\\
a^\bullet\circ\KK &=& \II & (a^\bullet~\mathit{comonoid~morphism})\\
\hline
\end{array}
$$

\begin{proposition}
An extensional $\SK$-algebra is equivalent to an
extensional $\BCIWK$-algebra. 
\end{proposition}
\begin{proposition}
For an extensional $\BCIWK$-algebra $\semcolor \calA$, $\semcolor \IA$ is a
closed cartesian operad 
s.t. $\semcolor \IA(0)\cong\calA$.
\end{proposition}
\begin{theorem}
Extensional $\BCIWK$-algebras are sound and complete with respect to
the  $\lambda\beta\eta$-calculus.
\end{theorem}

\begin{theorem}
The internal operad construction $\calA\mapsto \IA$ is
left adjoint to the functor from the category of closed cartesian operads to
that of extensional $\BCIWK$-algebras sending $\calP$ to $\calP(0)$.
\end{theorem}
This adjunction is actually  an {\em adjoint equivalence} 
(cf. the Fundamental Theorem  in \cite{Hyl17},
which covers non-extensional cases as well);
the cartesian case is technically much simpler than other variations.

\section{Conclusion and Future Work}
\label{sec:conclusion}

We proposed to use {\kc (semi-)closed operads} as
an appropriate framework for discussing combinatory completeness
of combinatory algebras.
As an alternative of polynomials, we introduced {\kc internal operads}
which make sense for extensional planar, linear, braided as well as 
classical combinatory algebras. 
Among them, the braided case was not covered by the conventional 
``polynomials as functions'' approach, and this fact prompted us to 
introduce internal operads.
In our study, the planar case is of particular importance, as it serves as the common foundation of all other cases. 

It is shown that the internal operad construction is
left adjoint to the forgetful functor from closed operads to extensional combinatory algebras.
In addition, the internal operad construction is useful for deriving extensionality axioms in a systematic, semantics-oriented way.

\subsection{Future Work}

There are several cases yet to be covered. 
It should be possible to study Tomita's {\kc bi-BDI-algebras} \cite{Tom22} within our framework; it is likely that they 
correspond to (semi-){\kc bi-closed}  planar operads.
Also it would be interesting to study combinatory algebras corresponding to the {\kc  tangled} (or knotted) $\lambda$-calculus briefly mentioned in \cite{Has22}. 
For a possible direction,
see the discussion on traced combinatory algebras below. 

Another important direction is to relax the limitations of our approach.
Firstly, we cannot handle applicative structures which are not combinatory complete.
For example, the extensional theory of $\BB$-terms of Ikebuchi and Nakano \cite{IN19}
is not covered --- for lack of the $\II$-combinator, it does not give rise to 
an operad. It would be nice if we could extend our framework to cover such cases.
Secondly, it is desirable to have a weak internal operad construction  for non-extensional combinatory algebras, which would give rise to semi-closed operads.

Finally, in this paper we did not consider partial algebras
nor relation to realizability. For that direction it would be useful to have
a framework generalizing both ours and Turing categories \cite{CH08}.

\subsection{Traced Combinatory Algebras}

The graph $\semcolor\mathbf{Tr}$ shown in Figure \ref{fig:Tr} does not correspond to a $\lambda$-term, but 
has interpretations in some $\BCpmI$-algebras,
e.g., those arising as a reflexive object 
in a ribbon category, including the crossed $G$-set  of 
$G$-labelled infinite binary trees \cite{Has22}.
\begin{figure}

\begin{minipage}[b]{0.45\linewidth}
\begin{center}\unitlength=.65pt
\begin{picture}(100,110)
\thicklines
\put(50,60){\circle{80}}
\put(50,20){\vector(0,-1){20}}
\put(47,100){\vector(-1,0){1}}
\put(30,25.5){\vector(2,-1){1}}
\put(70.5,25.5){\vector(2,1){1}}
\put(90,63){\vector(0,1){1}}
\put(75,80){\vector(1,0){1}}
\put(75,40){\vector(1,0){1}}
\qbezier(10,60)(10,60)(50,60)
\put(40,60){\vector(1,0){1}}
\qbezier(50,60)(60,40)(70,40)\qbezier(70,40)(85,40)(85,40)
\qbezier(50,60)(60,80)(70,80)\qbezier(70,80)(85,80)(85,80)
\put(50,20){\lamnode}
\put(10,60){\lamnode}
\put(85,80){\appnode}
\put(85,40){\appnode}
\put(50,60){\lamnode}
\end{picture}
\end{center}  
\caption{The trace combinator $\mathbf{Tr}$}
\label{fig:Tr}
\end{minipage}
\begin{minipage}[b]{0.45\linewidth}
\begin{center}\unitlength=.45pt
\begin{picture}(180,120)(10,-20)
\thicklines
\put(50,60){\circle{80}}
\put(100,0){\line(0,-1){20}}
\qbezier(10,60)(10,60)(50,60)
\qbezier(50,60)(60,40)(70,40)\qbezier(70,40)(85,40)(85,40)
\qbezier(50,60)(60,80)(70,80)\qbezier(70,80)(85,80)(85,80)
\put(150,60){\circle{80}}
\put(100,20){\oval(100,40)[b]}
\put(115,80){\line(1,0){70}}
\put(110,60){\line(1,0){80}}
\put(115,40){\line(1,0){70}}
\rgbfbox{135}{35}{30}{50}{1}{1}{0}{}%
\put(50,20){\tinylamnode}
\put(10,60){\tinylamnode}
\put(85,80){\tinyappnode}
\put(85,40){\tinyappnode}
\put(50,60){\tinylamnode}
\put(150,20){\tinylamnode}
\put(115,80){\tinylamnode}
\put(110,60){\tinylamnode}
\put(115,40){\tinylamnode}
\put(185,80){\tinyappnode}
\put(190,60){\tinyappnode}
\put(185,40){\tinyappnode}
\put(100,0){\tinyappnode}
\end{picture}
\begin{picture}(40,120)
\put(20,60){\makebox(0,0){$=$}}
\end{picture}
\begin{picture}(100,120)
\thicklines
\put(50,70){\circle{100}}
\put(10,100){\line(1,0){80}}
\put(2,80){\line(1,0){96}}
{\kc
\put(35,50){\oval(20,20)[l]}
\put(65,50){\oval(20,20)[r]}
\put(35,40){\line(1,0){30}}
}%
\rgbfbox{35}{55}{30}{50}{1}{1}{0}{}%
\put(50,20){\line(0,-1){20}}
\put(10,100){\tinylamnode}
\put(90,100){\tinyappnode}
\put(2,80){\tinylamnode}
\put(98,80){\tinyappnode}
\put(50,20){\tinylamnode}
\end{picture}
\end{center}
\caption{Applying $\mathbf{Tr}$}
\label{fig:Apply-Tr}
\end{minipage}

\end{figure}
By applying $\semcolor\mathbf{Tr}$, we can create {\em trace} \cite{JSV96}
in the internal PROP/PROB, as depicted in Figure \ref{fig:Apply-Tr}.
Such a trace operator allows us to represent knots and tangles. 
For instance, the trefoil knot can be expressed as the braid closure 
$
\mathbf{Tr}\,(\mathbf{Tr}\,(\CC^+\circ\CC^+\circ\CC^+))
$
of 
$ \CC^+\circ\CC^+\circ\CC^+$: 
\newsavebox{\boxTr}
\savebox{\boxTr}{%
\unitlength=.45pt
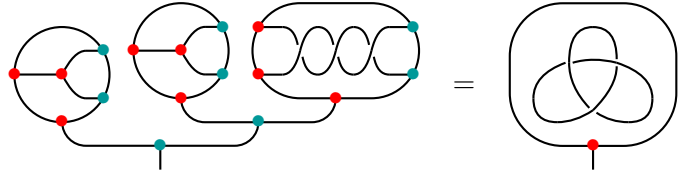
\begin{picture}(80,80)(10,20)\thicklines
\put(50,60){\circle{80}}
\qbezier(10,60)(10,60)(50,60)
\qbezier(50,60)(60,40)(70,40)\qbezier(70,40)(85,40)(85,40)
\qbezier(50,60)(60,80)(70,80)\qbezier(70,80)(85,80)(85,80)
\put(50,20){\tinylamnode}
\put(50,60){\tinylamnode}
\put(10,60){\tinylamnode}
\put(85,80){\tinyappnode}
\put(85,40){\tinyappnode}
\end{picture}
}%
\newsavebox{\boxbraid}
\savebox{\boxbraid}{%
\unitlength=.45pt
\begin{picture}(30,40)\thicklines
\qbezier(0,0)(10,0)(15,20)\qbezier(15,20)(20,40)(30,40)
\qbezier(0,40)(10,40)(13,25)\qbezier(17,15)(20,0)(30,0)
\end{picture}
}%
\begin{center}
\unitlength=.45pt
\begin{picture}(340,140)
\thicklines
%
\put(270,100){\oval(140,80)}
\qbezier(205,120)(205,120)(225,120)
\qbezier(205,80)(205,80)(225,80)
\put(225,80){\usebox{\boxbraid}}
\put(255,80){\usebox{\boxbraid}}
\put(285,80){\usebox{\boxbraid}}
\qbezier(315,120)(335,120)(335,120)
\qbezier(315,80)(335,80)(335,80)
\put(205,60){\oval(130,40)[b]}
\put(122.5,40){\oval(165,40)[b]}
\put(122.5,20){\line(0,-1){20}}
\put(0,40){\usebox{\boxTr}}
\put(100,60){\usebox{\boxTr}}
\put(270,60){\tinylamnode}
\put(205,120){\tinylamnode}
\put(205,80){\tinylamnode}
\put(335,120){\tinyappnode}
\put(335,80){\tinyappnode}
\put(205,40){\tinyappnode}
\put(122.5,20){\tinyappnode}
\end{picture}
\begin{picture}(60,140)
\put(30,70){\makebox(0,0){$=$}}
\end{picture}
\begin{picture}(140,140)
\thicklines
\put(70,80){\oval(140,120)}
\put(70,20){\line(0,-1){20}}
%
\qbezier(70,120)(50,120)(50,90)\qbezier(50,90)(50,70)(67,52)
\qbezier(73,48)(85,40)(100,40)\qbezier(100,40)(120,40)(120,60)
\qbezier(120,60)(120,80)(90,90)\qbezier(90,90)(70,95)(53,90)
\qbezier(47,90)(20,80)(20,60)\qbezier(20,60)(20,40)(40,40)
\qbezier(40,40)(55,40)(70,50)\qbezier(70,50)(90,70)(90,87)
\qbezier(90,93)(90,120)(70,120)
\put(70,20){\tinylamnode}
\end{picture}
\end{center}
Actually $\semcolor\mathbf{Tr}$ is far more expressive than one might expect.
With $\semcolor\mathbf{Tr}$, we can define the combinators 
$$\semcolor
\begin{array}{rcllcrcll}
\eta&=&\mathbf{Tr}\,(\mathbf{Tr}\circ(\BB\,\mathbf{Tr})\circ(\BB\,\CC)\circ\CC)
&:0\rightarrow2 & ~~\mathrm{and}~~ &
\varepsilon&=&\mathbf{Tr}\,(\CC\circ(\BB\,\CC)\circ(\BB\,\BB)\circ\BB)
&:2\rightarrow0
\end{array}
$$
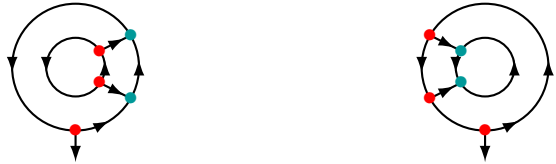
\begin{center}\unitlength=.6pt
\begin{picture}(100,80)
\thicklines
\put(50,60){\circle{80}}
\put(50,60){\circle{37}}
\put(50,20){\vector(0,-1){20}}
\put(70.5,25.5){\vector(2,1){1}}
\put(90,64){\vector(0,1){1}}
\put(68.5,64){\vector(0,1){1}}
\put(10,57){\vector(0,-1){1}}
\put(31.5,57){\vector(0,-1){1}}
\put(80,77.5){\vector(2,1){1}}
\put(80,42.5){\vector(2,-1){1}}
\qbezier(65,50)(85,40)(85,40)
\qbezier(65,70)(85,80)(85,80)
\put(65,70){\tinylamnode}
\put(65,50){\tinylamnode}
\put(85,80){\tinyappnode}
\put(85,40){\tinyappnode}
\put(50,20){\tinylamnode}
\end{picture}
~~~~~~~~~~~~~~~~~~~~~~~~
\begin{picture}(100,80)
\thicklines
\put(50,60){\circle{80}}
\put(50,60){\circle{37}}
\put(50,20){\vector(0,-1){20}}
\put(70.5,25.5){\vector(2,1){1}}
\put(90,64){\vector(0,1){1}}
\put(68.5,64){\vector(0,1){1}}
\put(10,57){\vector(0,-1){1}}
\put(31.5,57){\vector(0,-1){1}}
\put(30,72.5){\vector(2,-1){1}}
\put(30,47.5){\vector(2,1){1}}
\qbezier(35,50)(15,40)(15,40)
\qbezier(35,70)(15,80)(15,80)
\put(35,70){\tinyappnode}
\put(35,50){\tinyappnode}
\put(15,80){\tinylamnode}
\put(15,40){\tinylamnode}
\put(50,20){\tinylamnode}
\end{picture}
\end{center} 
which satisfy the zig-zag equation
$\semcolor\eta\circ(\BB\,\varepsilon)=(\BB\,\eta)\circ\varepsilon=\II$, e.g.
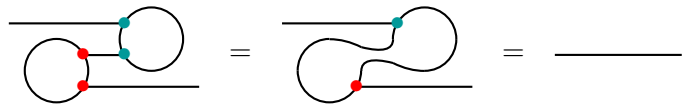
\begin{center}
\unitlength=.6pt
\begin{picture}(120,60)(40,30)
\thicklines
\qbezier(40,80)(100,80)(115,80)
\qbezier(85,60)(100,60)(115,60)
\qbezier(85,40)(100,40)(160,40)
\put(70,50){\oval(40,40)}
\put(130,70){\oval(40,40)}
\put(87,60){\tinylamnode}
\put(87,40){\tinylamnode}
\put(113,80){\tinyappnode}
\put(113,60){\tinyappnode}
\end{picture}
\begin{picture}(40,60)
\put(20,30){\makebox(0,0){$=$}}
\end{picture}
\begin{picture}(120,60)(40,30)
\thicklines
\qbezier(40,80)(100,80)(115,80)
\qbezier(85,40)(100,40)(160,40)
\qbezier(70,70)(80,70)(90,65)\qbezier(90,65)(110,60)(110,70)
\qbezier(90,50)(90,60)(110,55)\qbezier(110,55)(120,50)(130,50)
\put(70,50){\oval(40,40)[b]}\put(70,50){\oval(40,40)[tl]}
\put(130,70){\oval(40,40)[t]}\put(130,70){\oval(40,40)[br]}
\put(87,40){\tinylamnode}
\put(113,80){\tinyappnode}
\end{picture}
\begin{picture}(40,60)
\put(20,30){\makebox(0,0){$=$}}
\end{picture}
\begin{picture}(80,60)
\thicklines
\qbezier(0,30)(0,30)(80,30)
\end{picture}
\end{center}
Thus the internal PROP is not just traced but also {\kc compact closed} (ribbon in the braided case).

It is tempting to call such combinatory algebras with $\mathbf{Tr}$ 
{\em traced combinatory algebras}; to be more precise, a 
traced combinatory algebra should be an extensional
$\BCpmI$-algebra (or $\BCI$-algebra in the symmetric case)
equipped with a trace combinator $\mathbf{Tr}$.
The axiomatizations of traced combinatory algebras, and the corresponding
tangled $\lambda$-calculus, are left as an interesting future work.

\bibliographystyle{entics}

\appendix

\section{Braided Operads}
\label{sec:braided-operads}

\subsection{Braid Groups}
Let $B_n$ be the Artin braid group with $n$ strands \cite{Art25,KT08}, which can be 
represented by $n-1$ generators $\sigma_1$,\dots,$\sigma_{n-1}$
and the relations
$$
\sigma_i\sigma_j = \sigma_j\sigma_i ~~( j<i-1),~~~~
\sigma_i\sigma_{i+1}\sigma_i =\sigma_{i+1}\sigma_i\sigma_{i+1}
$$
Figure 
\ref{fig:relations} illustrates a graphical reading of the  
relations of $B_4$. 

Let $S_n$ be the symmetric group on $\{1,\dots,n\}$.
Let us denote the obvious homomorphism from $B_n$ to $S_n$ sending $\sigma_i$ to the permutation $(i,i+1)$ by $|\_|:B_n\rightarrow S_n$.
\begin{figure}[b!]
\unitlength=.8pt
$$
\begin{array}{rclrclrcl}
\sigma_3\sigma_1 &=& \sigma_1\sigma_3
&
\sigma_1\sigma_2\sigma_1 &=& \sigma_2\sigma_1\sigma_2
&
\sigma_2\sigma_3\sigma_2 &=& \sigma_3\sigma_2\sigma_3
\\
\\
\small
\begin{picture}(40,60)\thicklines
\put(0,0){\usebox{\hbraid}}
\myline{0}{40}{20}{40}
\myline{0}{60}{20}{60}
\put(20,40){\usebox{\hbraid}}
\myline{20}{0}{40}{0}
\myline{20}{20}{40}{20}
\put(3,-5){\dashbox(14,70){}}
\put(23,-5){\dashbox(14,70){}}
\put(10,-10){\makebox(0,0){$\sigma_1$}}
\put(30,-10){\makebox(0,0){$\sigma_3$}}
\end{picture}
&
\begin{picture}(10,60)
\put(5,30){\makebox(0,0){$=$}}
\end{picture}
&
\small
\begin{picture}(50,60)\thicklines
\put(0,40){\usebox{\hbraid}}
\myline{40}{40}{20}{40}
\myline{40}{60}{20}{60}
\put(20,0){\usebox{\hbraid}}
\myline{20}{0}{0}{0}
\myline{20}{20}{0}{20}
\put(3,-5){\dashbox(14,70){}}
\put(23,-5){\dashbox(14,70){}}
\put(10,-10){\makebox(0,0){$\sigma_3$}}
\put(30,-10){\makebox(0,0){$\sigma_1$}}
\end{picture}
~~~
&
~~~
\small
\begin{picture}(60,60)\thicklines
\put(0,0){\usebox{\hbraid}}  \myline{40}{40}{60}{40}
\myline{20}{40}{0}{40} \put(20,20){\usebox{\hbraid}}
\put(40,0){\usebox{\hbraid}} \myline{40}{0}{20}{0}
\myline{0}{60}{60}{60}
\put(3,-5){\dashbox(14,70){}}
\put(23,-5){\dashbox(14,70){}}
\put(43,-5){\dashbox(14,70){}}
\put(10,-10){\makebox(0,0){$\sigma_1$}}
\put(30,-10){\makebox(0,0){$\sigma_2$}}
\put(50,-10){\makebox(0,0){$\sigma_1$}}
\end{picture}
&
\begin{picture}(10,60)
\put(5,30){\makebox(0,0){$=$}}
\end{picture}
&
\small
\begin{picture}(60,60)\thicklines
\put(0,20){\usebox{\hbraid}}  \myline{40}{40}{20}{40}
\myline{20}{0}{0}{0} \put(20,0){\usebox{\hbraid}}
\put(40,20){\usebox{\hbraid}} \myline{40}{0}{60}{0}
\myline{0}{60}{60}{60}
\put(3,-5){\dashbox(14,70){}}
\put(23,-5){\dashbox(14,70){}}
\put(43,-5){\dashbox(14,70){}}
\put(10,-10){\makebox(0,0){$\sigma_2$}}
\put(30,-10){\makebox(0,0){$\sigma_1$}}
\put(50,-10){\makebox(0,0){$\sigma_2$}}
\end{picture}
~~~
&
~~~
\small
\begin{picture}(70,60)(-10,0)\thicklines
\put(0,20){\usebox{\hbraid}}  \myline{40}{60}{60}{60}
\myline{20}{60}{0}{60} \put(20,40){\usebox{\hbraid}}
\put(40,20){\usebox{\hbraid}} \myline{40}{20}{20}{20}
\myline{0}{0}{60}{0}
\put(3,-5){\dashbox(14,70){}}
\put(23,-5){\dashbox(14,70){}}
\put(43,-5){\dashbox(14,70){}}
\put(10,-10){\makebox(0,0){$\sigma_2$}}
\put(30,-10){\makebox(0,0){$\sigma_3$}}
\put(50,-10){\makebox(0,0){$\sigma_2$}}
\end{picture}
&
\begin{picture}(10,60)
\put(5,30){\makebox(0,0){$=$}}
\end{picture}
&
\small
\begin{picture}(60,60)\thicklines
\put(0,40){\usebox{\hbraid}}  \myline{40}{20}{60}{20}
\myline{20}{20}{0}{20} \put(20,20){\usebox{\hbraid}}
\put(40,40){\usebox{\hbraid}} \myline{40}{60}{20}{60}
\myline{0}{0}{60}{0}
\put(3,-5){\dashbox(14,70){}}
\put(23,-5){\dashbox(14,70){}}
\put(43,-5){\dashbox(14,70){}}
\put(10,-10){\makebox(0,0){$\sigma_3$}}
\put(30,-10){\makebox(0,0){$\sigma_2$}}
\put(50,-10){\makebox(0,0){$\sigma_3$}}
\end{picture}
\end{array}
$$
\caption{Axioms of $B_4$}
\label{fig:relations}
\end{figure}
For $s\in B_k$ and non-negative integers $j_1,\dots,j_k$, let
$s[j_1,\dots,j_k]\in B_{j_1+\dots+ j_k}$ be the braid obtained from
$s$ by replacing the $m$-th strand by $j_m$ parallel strands for $m=1,\dots,k$. 
When $j_m=0$, the $m$-th strand is simply deleted.
(For the extreme case that all $j_m$ are $0$, let $B_0$ be the trivial group.)
In our previous work on the braided $\lambda$-calculus \cite{Has22}, 
$s[j_1,\dots,j_k]\in B_{j_1+\dots j_k}$ amounts to
the substitution map 
$s[k:=j_k]\dots[1:=j_1]$.

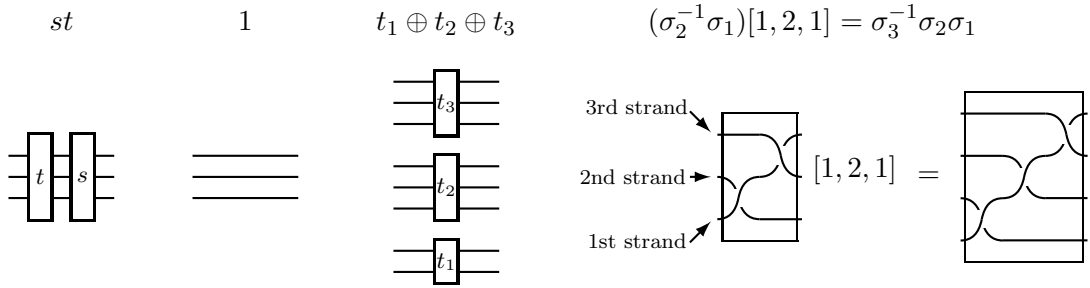
\begin{figure}
\unitlength=.8pt
$$
\begin{array}{cccc}
st
&
1
&
t_1\oplus t_2\oplus t_3
&
(\sigma_2^{-1}\sigma_1)[1,2,1] = \sigma_3^{-1}\sigma_2\sigma_1
\\
\\
\small
\begin{picture}(50,70)(0,-30)\thicklines
\myline{0}{10}{10}{10}
\myline{0}{20}{10}{20}
\myline{0}{30}{10}{30}
\put(10,0){\framebox(10,40){$t$}}
\myline{20}{10}{30}{10}
\myline{20}{20}{30}{20}
\myline{20}{30}{30}{30}
\put(30,0){\framebox(10,40){$s$}}
\myline{40}{10}{50}{10}
\myline{40}{20}{50}{20}
\myline{40}{30}{50}{30}
\end{picture}
&
~~~~~~~
\begin{picture}(50,70)(0,-30)\thicklines
\myline{0}{10}{50}{10}
\myline{0}{20}{50}{20}
\myline{0}{30}{50}{30}
\end{picture}
~~~~~~~
&
\small
\begin{picture}(50,90)\thicklines
\myline{0}{5}{20}{5}
\myline{0}{15}{20}{15}
\myline{0}{35}{20}{35}
\myline{0}{45}{20}{45}
\myline{0}{55}{20}{55}
\myline{0}{75}{20}{75}
\myline{0}{85}{20}{85}
\myline{0}{95}{20}{95}
\put(20,0){\framebox(10,20){$t_1$}}
\put(20,30){\framebox(10,30){$t_2$}}
\put(20,70){\framebox(10,30){$t_3$}}
\myline{30}{5}{50}{5}
\myline{30}{15}{50}{15}
\myline{30}{35}{50}{35}
\myline{30}{45}{50}{45}
\myline{30}{55}{50}{55}
\myline{30}{75}{50}{75}
\myline{30}{85}{50}{85}
\myline{30}{95}{50}{95}
\end{picture}
&
~~~~~~~~~~~
\begin{picture}(110,90)(-10,-10)
\put(32.5,10){\framebox(35,60){}}
\thicklines
\put(30,20){\usebox{\hbraid}}
\put(50,40){\usebox{\hbraidInv}}
\myline{30}{60}{50}{60}
\myline{50}{20}{70}{20}
\put(75,40){$[1,2,1]$}
\put(-8,10){\makebox(0,0){\scriptsize 1st strand}} \put(17,9){\vector(1,1){10}}
\put(-10,40){\makebox(0,0){\scriptsize  2nd strand}}  \put(17,40){\vector(1,0){10}}
\put(-8,73){\makebox(0,0){\scriptsize  3rd strand}} \put(17,71){\vector(1,-1){10}}
\end{picture}
\begin{picture}(50,90)
\put(30,50){\makebox(0,0){$=$}}
\end{picture}
\begin{picture}(60,90)(5,-10)
\put(2.5,0){\framebox(55,80){}}
\thicklines
\myline{0}{70}{40}{70}
\myline{0}{50}{20}{50}
\myline{40}{30}{60}{30}
\myline{20}{10}{60}{10}
\put(0,10){\usebox{\hbraid}}
\put(20,30){\usebox{\hbraid}}
\put(40,50){\usebox{\hbraidInv}}
\end{picture}
\end{array}
$$
\comment{
$$
\begin{array}{ccc}
\\
(\sigma_2^{-1}\sigma_1)[1,2,1] &=& \sigma_3^{-1}\sigma_2\sigma_1
\\
\\
\begin{picture}(110,80)
\put(32.5,10){\framebox(35,60){}}
\put(52.5,77.5){\makebox(0,0){$\sigma_2^{-1}\sigma_1$}}
\thicklines
\put(30,20){\usebox{\hbraid}}
\put(50,40){\usebox{\hbraidInv}}
\myline{30}{60}{50}{60}
\myline{50}{20}{70}{20}
\put(75,40){$[1,2,1]$}
\put(-8,10){\makebox(0,0){\footnotesize 1st strand}} \put(17,9){\vector(1,1){10}}
\put(-10,40){\makebox(0,0){\footnotesize  2nd strand}}  \put(17,40){\vector(1,0){10}}
\put(-8,73){\makebox(0,0){\footnotesize  3rd strand}} \put(17,71){\vector(1,-1){10}}
\end{picture}
&
\begin{picture}(40,80)
\put(20,40){\makebox(0,0){$=$}}
\end{picture}
&
\begin{picture}(60,80)
\put(2.5,0){\framebox(55,80){}}
\thicklines
\myline{0}{70}{40}{70}
\myline{0}{50}{20}{50}
\myline{40}{30}{60}{30}
\myline{20}{10}{60}{10}
\put(0,10){\usebox{\hbraid}}
\put(20,30){\usebox{\hbraid}}
\put(40,50){\usebox{\hbraidInv}}
\end{picture}
\end{array}
$$
}
\caption{Some constructions on braids}
\label{fig:constructions}
\end{figure}

For $t_1\in B_{j_1},\dots,t_k\in B_{j_k}$,
$t_1\oplus\dots\oplus t_k\in B_{j_1+\dots +j_k}$ is the block direct sum of the braids $t_1,\dots,t_k$. 
See Figure \ref{fig:constructions} for a graphical account of
these constructions.

\subsection{Braided Operads}
A {\em braided operad} \cite{Fie96}  
is an operad $\calP=(\calP(n))_{n\in\mathbf{N}}$
equipped with 
actions of the braid groups $\calP(j)\times B_j\rightarrow\calP(j)$
satisfying the following equivariance conditions
$$
\begin{array}{rcl}
(fs)(g_1,\dots,g_k) &=& (f(g_{s^{-1}(1)},\dots,g_{s^{-1}(k)}))s[j_1,\dots,j_k]\\
f((g_1t_1),\dots,(g_kt_k)) &=& (f(g_1,\dots,g_k))(t_1\oplus\dots\oplus t_k)
\end{array}
$$
where $f\in\calP(k)$ and $g_i\in \calP(j_i)$; $s\in B_k$ acts on $\{1,\dots,k\}$ via the homomorphism $|\_|:B_k\rightarrow S_k$
(so $s(i)=|s|(i)$). Figure \ref{fig:equivariance} in Section \ref{sec:variations}
illustrates an instance of the first equivariance condition.

\end{document}